\documentclass[showpacs,aps,twocolumn]{revtex4}
\usepackage{epsfig}
\usepackage{graphicx}
\usepackage{amsmath,amssymb,amsfonts}
\usepackage{array}
\usepackage{url}
\usepackage{hyperref}
\usepackage{multirow}
\usepackage{float}
\usepackage{lineno}
\usepackage{xspace}
\usepackage{caption}
\usepackage{subcaption}
\usepackage{ulem}
\usepackage{adjustbox}
\usepackage{csquotes}
\usepackage[usenames,dvipsnames]{color}

\newcommand{\bef}{\begin{figure}}
\newcommand{\eef}{\end{figure}}
\newcommand{\bc}{\begin{center}}
\newcommand{\ec}{\end{center}}

\newcommand{\be}{\begin{equation}}
\newcommand{\ee}{\end{equation}}
\newcommand{\bea}{\begin{eqnarray}}
\newcommand{\eea}{\end{eqnarray}}

\def\ba{\begin{eqnarray}}
\def\ea{\end{eqnarray}}
\begin{document}
\title{Effect of a magnetic field on the thermodynamic properties of a high-temperature hadron resonance gas with van der Waals interactions}
 
\author{Bhagyarathi Sahoo}
\author{Kshitish Kumar Pradhan}
\author{Dushmanta Sahu}
\author{Raghunath Sahoo\footnote{Corresponding author: Raghunath.Sahoo@cern.ch}}
\affiliation{Department of Physics, Indian Institute of Technology Indore, Simrol, Indore 453552, India}

\begin{abstract}
We study the behavior of a hadronic matter in the presence of an external magnetic field within the van der Waals hadron resonance gas model, considering both attractive and repulsive interactions among the hadrons. Various thermodynamic quantities like pressure ($P$), energy density ($\varepsilon$), magnetization ($\mathcal{M}$), entropy density ($s$), squared speed of sound  ($c_{\rm s}^{2}$), and specific-heat capacity at constant volume ($c_{v}$) are calculated as functions of temperature ($T$) and static finite magnetic field ($eB$). We also consider the effect of baryochemical potential ($\mu_{B}$) on the above-mentioned thermodynamic observables in the presence of a magnetic field. Further, we estimate the magnetic susceptibility ($\chi_{\rm M}^{2}$), relative permeability ($\mu_{\rm r}$), and electrical susceptibility ($\chi_{\rm Q}^{2}$) which can help us to understand the system better. Through this model, we quantify a liquid-gas phase transition in the T-eB-$\mu_B$ phase space.

\end{abstract}
\date{\today}
\maketitle

\section{Introduction}
\label{intro}

In the early stages of the evolution of the Universe, it was supposed to be extremely hot and dense, possibly filled with a unique state of matter called quark-gluon plasma (QGP). We explore the ultrarelativistic heavy-ion collisions in laboratories to probe such initial conditions. At extreme temperatures and/or baryon densities, the hadronic degrees of freedom transform into partonic degrees of freedom, resulting in QGP formation. Quantum chromodynamics (QCD) is the widely used theory to describe the behavior of QGP. In addition, studying its thermodynamic properties is of utmost importance to understand the behavior and evolution of hot and dense QCD matter. Various thermodynamic properties of strongly interacting nuclear matter have been estimated from the first-principle lattice QCD (lQCD) approach. However, the applicability of lQCD breaks down at high baryochemical potential due to the fermion sign problem~\cite{Borsanyi:2013bia, HotQCD:2014kol}. An alternative to the lQCD approach at low temperatures (up to 150 MeV) is the hadron resonance gas (HRG) model. The HRG model has been observed to agree with the lQCD results for temperatures up to $T \simeq 140-150$ MeV at zero baryochemical potential~\cite{Bellwied:2013cta, HotQCD:2012fhj, Bellwied:2017ttj, Borsanyi:2010cj, Karsch:2003vd, Huovinen:2009yb}. The HRG model is thus a better alternative to study the baryon-rich environments at low-temperature regimes~\cite{Karsch:2003zq, Tawfik:2004sw, Allton:2005gk, Borsanyi:2012cr}.

In an ideal HRG model~\cite{Braun:2004, Dashen:1969ep, Venugopalan:1992hy, Bhattacharyya:2015pra}, the hadrons are assumed to be pointlike particles with no interaction between them. However, this assumption is very simplistic and fails to describe the lQCD data at temperatures above $T \simeq$ 150 MeV, where the hadrons melt down and the HRG model reaches its limits. Although this shortcoming of the HRG model can be easily ignored while studying the thermodynamic properties, however, while estimating various charge fluctuations at higher order, the shortcomings of the HRG model are not trivial. Recently, much focus has been diverting towards an interacting hadron resonance gas model as they extend the region of agreement with lQCD data due to the interactions between the hadrons. Excluded-volume hadron resonance gas (EVHRG) model assumes an eigenvolume parameter for the hadrons, which essentially mimics a repulsive interaction in the hadron gas ~\cite{Hagedorn:1980kb, Rischke:1991ke, Cleymans:1992jz, Yen:1997rv, Fu:2012zzc, Fu:2013gga, Andronic:2012ut, Singh:1991np, Gorenstein:2007mw, Begun:2012rf, Tawfik:2013eua}. Unequal sizes of different hadron species are handled by modified excluded-volume hadron resonance gas model~\cite{Bhattacharyya:2013oya, Bugaev:2000wz, Sagun:2013moa}. Similarly, the mean-field hadron resonance gas model introduces a repulsive interaction potential in the hadronic medium~\cite{Pal:2021qav, Kadam:2019peo, Anchishkin:2014hfa}. There are also various other improvements to the HRG model in literature, such as the Lorentz modified excluded-volume hadron resonance gas model ~\cite{Pal:2020ink}, where the hadrons are treated as Lorentz-contracted particles, and the effective thermal mass hadron resonance gas model ~\cite{Zhang:2019uct}, where the hadrons gain effective mass with temperature. However, the most successful improvement to the model which explains the lQCD results is the van der Waals hadron resonance gas (VDWHRG) model~\cite{Samanta:2017yhh,Vovchenko:2015vxa, Vovchenko:2016rkn, Sarkar:2018mbk,Pradhan:2022gbm}. This model assumes a van der Waals-type interaction between the hadrons, having both attractive and repulsive parts. The VDWHRG model effectively explains the lQCD data up to $T \simeq 180$ MeV. From this, we can infer that van der Waals interaction does play a crucial role in the hadronic systems at high temperatures. Moreover, the VDWHRG model has been recently used to estimate various thermodynamic and transport properties~\cite{Samanta:2017yhh,Pradhan:2022gbm,Sarkar:2018mbk}, along with fluctuations of conserved charges~\cite{Samanta:2017yhh}, which show a good agreement with the lQCD estimations. In addition, there are several studies exploring the liquid-gas phase transition using the VDWHRG model, locating a possible critical point for the phase transition \cite{Samanta:2017yhh, Vovchenko:2015vxa, Sarkar:2018mbk}

A unique consequence of the peripheral heavy-ion collisions is that a strong transient magnetic field ( $\sim m_{\pi}^{2} \sim 10^{18} $ G) is expected to be formed due to the motion of the spectator protons. The strength of the magnetic field may reach up to the order of 0.1$m_{\pi}^{2}$, $m_{\pi}^{2}$, $15 m_{\pi}^{2}$ for Super Proton Synchroton (SPS), Relativistic Heavy Ion Collider (RHIC), and LHC energies, respectively~\cite{Skokov:2009qp}. This magnetic field decays with time and can, in principle, affect the thermodynamic and transport properties of the evolving partonic and hadronic matter~\cite{Skokov:2009qp, Tawfik:2016lih, Bzdak:2011yy, Deng:2012pc}. The strong magnetic field, which can reach hadronic scales, has a significant effect on the transition properties and equation of state. Such intense magnetic fields are predicted to occur in compact neutron stars~\cite{Duncan:1992hi, Dey:2002pc} and during the early Universe's electroweak transition~\cite{Vachaspati:1991nm,Bhatt:2015ewa}. The interaction between the strong dynamics and the external magnetic field leads to exciting new phenomena, such as the chiral magnetic effect~\cite{Kharzeev:2007jp, Fukushima:2008xe} and a reduction of the transition temperature as the magnetic field increases~\cite{Bali:2011qj}. Furthermore, magnetic catalysis~\cite{Shovkovy:2012zn} and inverse magnetic catalysis ~\cite{Preis:2010cq,Bruckmann:2013oba} can affect the phase diagram of QCD matter. Thus, it is crucial to study the effect of an external magnetic field on both the deconfined and confined phases of the matter formed in high-energy collisions. Thermodynamic properties of the system, such as pressure ($P$), energy density ($\varepsilon$), entropy density ($s$), speed of sound ($c_{\rm s}$), and specific heat capacity at constant volume ($c_{\rm v}$) will get modified due to the effect of an external magnetic field. All these observables help us characterize the systems produced in ultrarelativistic collisions. Moreover, the system will also develop some magnetization ($\mathcal{M}$), which will help us to understand whether the system is diamagnetic or paramagnetic. Apart from these, the magnetic susceptibility ($\chi_{M}^{2}$) and magnetic permeability ($\mu_{r}$) are also essential observables that can give us useful information about the system under consideration~\cite{Bali:2012jv,Bali:2020bcn,Lin:2022ied, Xu:2020yag,Bonati:2013vba, Bonati:2013lca,Bali:2014kia,Chaudhuri:2022oru, Bhattacharyya:2015pra}. Thus, one must study the above-mentioned observables to better understand the nature and behavior of both the hadronic and partonic media formed in peripheral heavy-ion collisions.

Several works in literature concern with the study of the matter formed in ultrarelativistic collisions in the presence of a constant external magnetic field. In Ref. \cite{Bali:2014kia}, a detailed analysis of the hot and dense QCD matter in the presence of an external magnetic field has been done with the lQCD approach. The results from the $SU$(3) Polyakov linear-sigma model have also been contrasted with the existing lQCD estimations \cite{Tawfik:2016gye}. In addition, in Refs.~\cite{Endrodi:2013cs,Kadam:2019rzo} the authors use the HRG and EVHRG models in the presence of constant external magnetic fields to estimate the fundamental thermodynamic quantities such as pressure, energy density, and magnetization. Moreover, in Ref. \cite{Bhattacharyya:2015pra}, the authors discuss the effect of external magnetic field on the correlations and fluctuations of the hadron gas. An interesting study has been conducted by assuming an away-from-equilibrium scenario by employing the nonextensive Tsallis statistics, and then the basic thermodynamic quantities have been estimated~\cite{Pradhan:2021vtp}. In the present study, we use the van der Waals hadron resonance gas model, an improved and new approach to studying the hadronic medium of high-energy collisions. Furthermore, van der Waals interaction leads to a liquid-gas phase transition in the system along with a critical point. We can take advantage of this fact and study the QCD phase diagram. In literature, the QCD phase transition in the $T-\mu_{B}$ plane has been studied extensively from various models, including the VDWHRG model~\cite{Samanta:2017yhh,Sarkar:2018mbk}. A similar QCD phase transition in the $T-eB$ plane is also important to understand the QCD matter and its consequences. There are few studies where the authors have used various models to map the phase diagram. This study uses the hadron gas with van der Waals interaction and explores the possible critical point in the $T-eB-\mu_{B}$ plane. This paper is organized as follows. Section \ref{formulation} gives a detailed calculation of the thermodynamic observables and susceptibilities within the ambit of a VDWHRG model under an external magnetic field. In Sec, \ref{vacpressure}, we give the detailed calculation of the vacuum contribution to the thermodynamic observables due to the external magnetic field. We discuss the results in Sec. \ref{res} and briefly summarize our work in Sec.\ref{sum}.

\section{Formulation}
\label{formulation}

The ideal HRG formalism considers hadrons to be point particles with no interactions between them. Under this formalism, the partition function of $i$th particle species in a grand canonical ensemble (GCE) is given as ~\cite{Andronic:2012ut}
\begin{equation}
\label{eq1}
ln Z^{id}_i = \pm Vg_i\int \frac{d^3p}{(2\pi)^3}  \ ln\{1\pm \exp[-(E_i-\mu_i)/T]\},
\end{equation}
where $T$ is the temperature of the system and $V$ represents the volume. The notations $g_i$, $E_i = \sqrt{p^2 + m_i^2}$, $ m_i $ and $\mu_i$ are for the degeneracy, energy, mass, and chemical potential of the $i$th hadron, respectively. Here, $id$ refers to the ideal. The plus and minus signs ($\pm$) correspond to baryons and mesons, respectively. $\mu_i$ is further expanded in terms of the baryonic, strangeness, and charge chemical potentials ($\mu_{B}$, $\mu_{S}$ and $\mu_{Q}$, respectively) and the corresponding conserved numbers ($B_i$, $S_i$ and $Q_i$) as,
\begin{equation}
\label{eq2}
\mu_i = B_i\mu_B + S_i\mu_S +Q_i\mu_Q.
\end{equation}
The total grand canonical partition function of noninteracting hadron resonance gas is the sum of partition functions of all hadrons and resonances~\cite{Andronic:2012ut, Braun:2004},
\begin{equation}
\label{eq3}
ln Z^{id} = \sum_i ln Z^{id}_i.
\end{equation}
The free-energy density of the ideal HRG model can be written in terms of partition function  as,
\begin{equation}
\label{eq4}
    f^{id} = -T lnZ^{id}.
\end{equation}
 The ideal pressure is defined as the negative of free-energy density,
 \begin{equation}
 \label{eq5}
      P^{id} = - f^{id}.
 \end{equation}
 The explicit form of thermodynamic pressure $P_i$, energy density $\varepsilon_i$, number density $n_i$, and entropy density $s_i$ in the ideal HRG formalism can now be obtained as
\begin{equation}
\label{eq6}
P^{id}_i(T,\mu_i) = \pm Tg_i \int \frac{d^3p}{(2\pi)^3} \ ln\{1\pm \exp[-(E_i-\mu_i)/T]\}
\end{equation}
\begin{equation}
\label{eq7}
\varepsilon^{id}_i(T,\mu_i) = g_i \int \frac{ d^3p}{(2\pi)^3}  \frac{E_i \ }{\exp[(E_i-\mu_i)/T]\pm1}
\end{equation}
\begin{equation}
\label{eq8}
n^{id}_i(T,\mu_i) = g_i \int \frac{d^3p }{(2\pi)^3}\frac{1}{\exp[(E_i-\mu_i)/T]\pm1}
\end{equation}
\begin{align}
 s^{id}_i(T,\mu_i)=&\pm g_i \int \frac{d^3p }{(2\pi)^3} \Big[\ln\{1\pm  \exp[-(E_i-\mu_i)/T]\}\nonumber\\ 
&\pm \frac{(E_i-\mu_i)/T}{\exp[(E_i-\mu_i)/T]\pm 1}\Big].
 \label{eq9}
 \end{align}
 
  In the presence of a magnetic field ( for simplicity, suppose the magnetic field is pointing along $z$ direction), the single-particle energy for the charged and neutral particles is given as ~\cite{Landau, Endrodi:2013cs, Kadam:2019rzo}
 \begin{align}
\label{eq10}
  E^z_{c,i}(p_z,k,s_z) = \sqrt{p_z^2 + m_i^2+2|Q_i|B\bigg(k+\frac{1}{2}-s_z\bigg)}, Q_{i} \neq 0
\end{align}
 \begin{equation}
\label{eq11}
  E_{n,i}(p) = \sqrt{p^2 + m_i^2}, \hspace{0.5cm} Q_{i} = 0,
\end{equation}

where $Q_i$ is the charge of the $ith$ particle and $s_z$ is the component of spin $s$ in the direction of magnetic field $B$ and $k$ is the Landau level. The subscripts \enquote{$c$} and \enquote{$n$} are for charged and neutral particles.

In the presence of Landau level, one writes the three-dimensional integral as one-dimensional integral~\cite{Chakrabarty:1996te, Fraga:2008qn},
\begin{equation}
    \int \frac{d^3p}{(2\pi)^3} = \frac{|Q|B}{2\pi^{2}} \sum_{k} \sum_{s_z}   \int_{0}^{\infty} dp_z.
    \label{eq12}
\end{equation}

Now, in the presence of a finite magnetic field, the free energy of the system can be written as ~\cite{Kittel, Stanley}
 \begin{equation}
     f = \varepsilon - Ts - QB. \mathcal{M} ,
     \label{eq13}
 \end{equation}
  where, $\mathcal{M}$ is the magnetization. Further, in the presence of finite baryochemical potential, the above equation becomes
  \begin{equation}
     f = \varepsilon - Ts - QB. \mathcal{M} - \mu n ,      
  \end{equation}
  The $n$ is the number density. The above equation satisfies the differential relations,
 \begin{equation}
      s = - \frac{\partial f}{\partial T} ,  \hspace{1cm}  \mathcal{M} = - \frac{\partial f}{\partial (QB)},  \hspace{1cm} n = - \frac{\partial f}{\partial \mu}.
      \label{eq14}
 \end{equation}
In general, the free-energy density of the system contains contributions from both thermal and vacuum parts.
 \begin{equation}
     f  = f_{vac} +f_{th},
     \label{eq15}
 \end{equation}

 $f_{vac}$ and $f_{th}$ are the vacuum and thermal part of free-energy density, respectively. $f_{vac}$ is defined as the free-energy density at zero temperature and finite magnetic field, and $f_{th}$ is the free-energy at finite temperature and finite magnetic field.

 Furthermore, the total free-energy density of the HRG model in the presence of a magnetic field is due to the sum of independent contributions coming from all the hadrons $i$~\cite{Endrodi:2013cs},

 \begin{equation}
 \label{freeenergy}
     f = \sum_{i} g_i . f_{i}(\{T, Q_iB\},\{m_{i}, Q_{i}/e,s_{zi},\gamma_{i}\}).
 \end{equation}
 Here, $\gamma_{i}$ is the gyromagnetic ratio. To avoid the large uncertainty in the experimental values of gyromagnetic ratios (except for proton and neutron), it is chosen to be $\gamma_i$ = 2$Q_{i}$/e, as determined by the universal tree-level arguments~\cite{Ferrara:1992yc}. It is based upon the assumption that the considered hadrons are pointlike objects and all the neutral hadrons have a gyromagnetic ratio $\gamma_{i}$ = 0. To consider the accurate gyromagnetic ratios, an improvement method based on the generalized description of the anomalous magnetic moments for spin-1/2 and spin-1 particles has been developed in Refs.~\cite{Goldman:1972vc, Tsai:1971zma}. In Ref.~\cite{Tsai:1971zma}, it is found that in the presence of a homogeneous magnetic field, the coupling of anomalous magnetic moments with the spin-1 particles results in complex energy eigenvalues. This is due to the fact that for spin-1 theory, the anomalous magnetic moment does not vanish in a very strong magnetic field for the constant value of the anomalous magnetic moment. Thus the theory becomes inconsistent. The conditions for a consistent theory are discussed in Ref~\cite{Goldman:1972vc}. However, there are no such limitations for spin-1/2 particles. 
 
Now, the thermal part of the thermodynamic pressure, energy density, number density, and entropy density, i.e., Eqs.~(\ref{eq6}), (\ref{eq7}), (\ref{eq8}), and (\ref{eq9}) for charged particles in the presence of magnetic field can be modified using Eq.~(\ref{eq12}),

\begin{align}
\label{eq16}
P^{id,z}_{c,i}(T,\mu_i,B) = \pm \frac{Tg_i|Q_i|B}{2\pi^2} \sum_{k} \sum_{sz} \int_{0}^{\infty} dp_z \ ln\{1\pm \nonumber\\ \exp[-(E_{c,i}^{z}-\mu_i)/T]\}
\end{align}
\begin{align}
    \label{eq17}
\varepsilon^{id,z}_{c,i}(T,\mu_i,B) = \frac{g_i|Q_i|B}{2\pi^2}  \sum_{k} \sum_{sz} \int dp_z  E_{c,i}^{z}  \nonumber\\ \left[ \frac{1}{\exp[(E_{c,i}^{z}-\mu_i)/T]\pm1} \right]
\end{align}

\begin{align}
\label{eq18}
n^{id,z}_{c,i}(T,\mu_i,B) = \frac{g_i|Q_i|B}{2\pi^2}   \sum_{k} \sum_{sz} \int dp_z \  \nonumber\\ \left[ \frac{1}{\exp[(E_{c,i}^{z}-\mu_i)/T]\pm1} \right]
\end{align}
\begin{align}
 s^{id,z}_{c,i}(T,\mu_i,B)=&\pm \frac{g_i |Q_i|B}{2\pi^2}   \sum_{k} \sum_{sz} \int dp_z \Big[\ln\{1\pm \nonumber\\ \exp[-(E_{c,i}^z-\mu_i)/T]\} 
&\pm \frac{(E_{c,i}^{z}-\mu_i)/T}{\exp[(E_{c,i}^{z}-\mu_i)/T]\pm 1}\Big].
 \label{eq19}
 \end{align}
 
  For neutral particles, these thermodynamic variables are calculated using Eqs.~(\ref{eq6}), (\ref{eq7}), (\ref{eq8}), and (\ref{eq9}). The total pressure, energy density, and entropy density of the system are due to the sum of contributions from the charged particles and neutral particles. Now, we can use the above basic thermodynamic quantities to estimate other important observables.

The specific heat of the system is defined as the thermal variation of energy density at constant volume. It is defined as
 \begin{equation}
     c_{v} = \left(\frac{\partial \varepsilon}{\partial T}\right)_V.
     \label{eq20}
 \end{equation}

 The squared speed of sound is defined as the change in pressure of a system as a function of a change in energy density at constant entropy density per number density, i.e., $s/n$. Mathematically, the adiabatic squared speed of sound is defined as

 \begin{equation}
 c_{s}^{2} = \left(\frac{\partial P}{\partial \varepsilon}\right)_{s/n} = \frac{s}{c_{v}}.
     \label{eq21}
 \end{equation}

In the presence of both magnetic field and chemical potential, the squared speed of sound ($c_{s}^2$) is defined as,
\begin{equation}
c_{s}^{2}(T, \mu, QB) = \frac{\frac{\partial P}{\partial T} + \frac{\partial P}{\partial \mu} \frac{\partial \mu
}{\partial T} + \frac{\partial P}{\partial (QB)} \frac{\partial (QB
)}{\partial T}}{\frac{\partial \varepsilon}{\partial T} + \frac{\partial \varepsilon}{\partial \mu} \frac{\partial \mu}{\partial T} + \frac{\partial \varepsilon}{\partial (QB)} \frac{\partial (QB)}{\partial T}}
\label{21.1}
\end{equation}

where, 
\begin{equation}
    \frac{\partial (QB)}{\partial T} = \frac{ s\frac{\partial n}{\partial T} - n\frac{\partial s}{\partial T}}{n\frac{\partial s}{\partial (QB)} - s \frac{\partial n}{\partial (QB)} }
    \label{21.2}
\end{equation}
and,
\begin{equation}
    \frac{\partial \mu}{\partial T} = \frac{ s\frac{\partial n}{\partial T} - n\frac{\partial s}{\partial T}}{n\frac{\partial s}{\partial \mu} - s \frac{\partial n}{\partial \mu} }.
    \label{21.2}
\end{equation}

A detailed derivation of the squared speed of sound in the presence of a finite baryochemical potential and an external magnetic field is given in Appendix~\ref{appe1}.

The magnetization of the system can also be obtained from the following equation
\begin{equation}
    \mathcal{M} =\frac{\varepsilon_{tot} - \varepsilon}{QB},
    \label{eq}
\end{equation}
where, $ \varepsilon_{tot}$= $\varepsilon_{c, i}^{z} + \varepsilon_{n, i}$ is the energy density of the system in the presence of the magnetic field. $\varepsilon_{c, i}^{z}$, and $\varepsilon_{n, i} $ are the energy density of charged and neutral particles in the presence of a magnetic field, respectively. $\varepsilon $ is the free-energy density in the absence of a magnetic field.

We now proceed toward the estimation of the optical properties of a hadronic system. The derivative of magnetization with respect to the magnetic field is called magnetic susceptibility and is given by,

\begin{equation}
    \chi_{M}^{2} = \frac{\partial \mathcal{M}}{\partial (QB)} = \frac{\partial^{2} P }{\partial (QB)^{2}}.
    \label{eq22}
\end{equation}

From heavy-ion collision (HIC) perspectives, fluctuations of conserved charges have comparable importance as magnetic susceptibility since they play a vital role in describing QCD phase transition. The nth-order susceptibility is defined as
\begin{equation}
    \chi_{B/Q/S}^{n} = \frac{\partial^{n} (\frac{P}{T^{4}} )}{\partial (\frac{\mu_{B/Q/S}}{T})^{n}}.
    \label{eq22.1}
\end{equation}

 The second-order susceptibility corresponding to the electric charge chemical potential is called electric charge susceptibility, and is given by

\begin{equation}
    \chi_{Q}^{2} = \frac{1}{T^{2}}\frac{\partial^{2} P }{\partial \mu_{Q}^{2}}
    \label{eq23}
\end{equation}

The explicit forms of $ \chi_{M}^{2} $ and $\chi_{Q}^{2}$ are shown in Appendix ~\ref{appe2} and ~\ref{appe3}, respectively. 

 To include interactions in the hadronic system, we take advantage of the van der Waals equation of state. The ideal HRG model can be modified to include van der Waals interactions between particles by the introduction of the attractive and repulsive parameters $a$ and $b$, respectively. This modifies the pressure and number density obtained in ideal HRG iteratively as follows \cite{Samanta:2017yhh,Vovchenko:2015vxa,Vovchenko:2015pya} 
\begin{equation}
\label{eq24}
    P(T,\mu) = P^{id}(T,\mu^{*}) - an^{2}(T,\mu),
\end{equation}
where, the $n(T,\mu)$ is the VDW particle number density given by
\begin{equation}
\label{eq25}
    n(T,\mu) = \frac{\sum_{i}n_{i}^{id}(T,\mu^{*})}{1+b\sum_{i}n_{i}^{id}(T,\mu^{*})}.
\end{equation}
Here, $i$ runs over all hadrons and $\mu^{*}$ is the modified chemical potential given by,
\begin{equation}
\label{eq26}
    \mu^{*} = \mu - bP(T,\mu) - abn^{2}(T,\mu) + 2an(T,\mu).
\end{equation}
The transcendental equation in Eq. (\ref{eq24})-Eq. (\ref{eq26}) can be solved numerically. Starting with a given $\mu$ value, one can obtain ideal pressure and ideal number density given by Eq. (\ref{eq6}) and Eq. (\ref{eq8}). Using this, the VDW pressure and number density are calculated using Eq. (\ref{eq24}) and Eq. (\ref{eq25}) to finally obtain $\mu^*$. This $\mu^*$ is now used to calculate again the ideal pressure and number density, and the same process is repeated until the value of $\mu^*$ converges. Then Eq. (\ref{eq25}) gives the final VDW pressure, which can be used to obtain other thermodynamical quantities.

It is to be noted that the repulsive parameter is usually attributed to be related to the hard-core radius of the particle, $r$, by the relation $b = 16\pi r^{3}/3$. At the same time, the VDW parameter, $a$, represents the attractive interaction at an intermediate range. 

The entropy density $s(T,\mu)$ and energy density $\varepsilon(T,\mu)$ in VDWHRG can now be obtained as,
\begin{equation}
\label{eq27}
s(T,\mu) = \frac{s^{id}(T,\mu^{*})}{1+bn^{id}(T,\mu^{*})}
\end{equation}
\begin{equation}
\label{eq28}
\varepsilon(T,\mu) = \frac{\sum_{i}\varepsilon_{i}^{id}(T,\mu^{*})}{1+b\sum_{i}n_{i}^{id}(T,\mu^{*})} - an^{2}(T,\mu)
\end{equation}

The initial form of VDWHRG excluded interactions between baryon-antibaryon pairs and in between pairs involving at least one meson \cite{Samanta:2017yhh,Vovchenko:2015vxa,Vovchenko:2015pya,Vovchenko:2016rkn}. The baryon-antibaryon interactions were ignored under the assumption that annihilation processes dominate \cite{Andronic:2012ut,Vovchenko:2016rkn}. Meson interactions were ignored as their inclusion led to a suppression of thermodynamic quantities and could not explain the lQCD data at vanishing $\mu_B$ towards high temperatures \cite{Vovchenko:2016rkn}. The attractive and repulsive parameters, in this case, were derived either from properties of the ground state of nuclear matter \cite{Vovchenko:2015vxa} or by fitting the lQCD results for different thermodynamic quantities \cite{Samanta:2017yhh,Sarkar:2018mbk}. A formalism including the effect of meson-meson interactions through a hard-core repulsive radius ($r_M$) ~\cite{Sarkar:2018mbk} was developed where a simultaneous fit to the lQCD values was done to obtain the values of $a$ and $b$. The VDW parameters were considered to be fixed for all values of $\mu_B$ and $T$ in each of these implementations. The total pressure in the VDWHRG model is then written as \cite{Samanta:2017yhh,Vovchenko:2015vxa,Vovchenko:2015pya,Vovchenko:2016rkn,Sarkar:2018mbk},

\begin{equation}
\label{finalp}
P(T,\mu) = P_{M}(T,\mu) + P_{B}(T,\mu) + P_{\bar{B}}(T,\mu).
\end{equation}
Here, the $P_{M}(T,\mu), P_{B(\bar B)}(T,\mu)$ are the contributions to pressure from mesons and (anti)baryons, respectively, and are given by
\begin{equation}
\label{mesonp}
P_{M}(T,\mu) = \sum_{i\in M}P_{i}^{id}(T,\mu^{*M}),       
\end{equation}
\begin{equation}
\label{baryonp}
P_{B}(T,\mu) = \sum_{i\in B}P_{i}^{id}(T,\mu^{*B})-an^{2}_{B}(T,\mu),
\end{equation}
\begin{equation}
\label{antip}
P_{\bar{B}}(T,\mu) = \sum_{i\in \bar{B}}P_{i}^{id}(T,\mu^{*\bar{B}})-an^{2}_{\bar{B}}(T,\mu).
\end{equation}
Here, $M$, $B$, and $\bar B$ represent mesons, baryons, and antibaryons, respectively. $\mu^{*M}$ is the modified chemical potential of mesons because of the excluded volume correction, and $\mu^{*B}$ and $\mu^{*\bar B}$ are the modified chemical potentials of baryons and antibaryons due to VDW interactions~\cite{Sarkar:2018mbk}. Considering the simple case of vanishing electric charge and strangeness chemical potentials, $\mu_{Q} = \mu_{S} = 0$, the modified chemical potential for mesons and (anti)baryons can be obtained from Eq.~(\ref{eq2}) and Eq.~(\ref{eq27}) as; 
\begin{equation}
\label{mumeson}
\mu^{*M} = -bP_{M}(T,\mu),
\end{equation}
\begin{equation}
\label{mubaryon}
\mu^{*B(\bar B)} = \mu_{B(\bar B)}-bP_{B(\bar B)}(T,\mu)-abn^{2}_{B(\bar B)}+2an_{B(\bar B)},
\end{equation}
where $n_{M}$, $n_{B}$ and $n_{\bar B}$ are the modified number densities of mesons, baryons, and antibaryons, respectively, which are given by,
\begin{equation}
\label{nmeson}
    n_{M}(T,\mu) = \frac{\sum_{i\in M}n_{i}^{id}(T,\mu^{*M})}{1+b\sum_{i\in M}n_{i}^{id}(T,\mu^{*M})},
\end{equation}
\begin{equation}
\label{nbaryon}
    n_{B(\bar B)}(T,\mu) = \frac{\sum_{i\in B(\bar B)}n_{i}^{id}(T,\mu^{*B(\bar B)})}{1+b\sum_{i\in B(\bar B)}n_{i}^{id}(T,\mu^{*B(\bar B)})}.
\end{equation}

For this work, the parameters in the model are taken as $a=0.926$ GeV fm$^{3}$ and $b=(16/3)\pi r^3$, where the hard-core radius $r$ is replaced by $r_{M}=0.2$ fm and $r_{B,(\bar{B})}=0.62$ fm, respectively for mesons and (anti)baryons~\cite{Sarkar:2018mbk}. Now, we take the magnetic field-modified total ideal pressure, energy density, and entropy density and use them in the respective VDW equations to estimate the required thermodynamic observables.

\section{Renormalization of vacuum pressure}
\label{vacpressure}

As we discussed in the previous section, the total pressure (negative of the total free-energy density) of the system is due to both the thermal and vacuum components, i.e. 
\begin{equation}
    P_{total} = P_{th} (T, eB) + \Delta P_{vac}(T=0, eB)
\end{equation}
where $P_{th} (T, eB)$ is the thermal part of the pressure, which is the sum of the pressure due to both charged and neutral particles. In the presence of a magnetic field, the thermal parts of the pressure for charged and neutral particles are calculated using Eqs.~(\ref{eq16}), and~(\ref{eq6}), respectively.
 In this section, we will calculate the vacuum contribution of pressure in the presence of an external magnetic field using a dimensional regularization method. The vacuum pressure term is ultraviolet divergent, and it requires appropriate regularization to extract meaningful physical information\cite{Kadam:2019rzo,Endrodi:2013cs,Menezes:2008qt}. As a result, magnetic field-dependent and independent components must be distinguished using an appropriate regularization technique.

 In the presence of an external magnetic field, the vacuum pressure  for a charged spin-$\frac{1}{2}$ particle is given by \cite{Kadam:2019rzo,Endrodi:2013cs,Menezes:2008qt}
 
\begin{equation}
\label{eq29}
P_{\text{vac}}(S=1/2,B)= \frac{1}{2}\sum_{k=0}^{\infty}g_{k}\frac{|Q|B}{2\pi}\int_{-\infty}^{\infty}\frac{dp_z}{2\pi}E_{p,k}(B),
\end{equation}
where $g_{k}=2-\delta_{k0}$ is the degeneracy of the $kth$ Landau level. We have added and subtracted the lowest Landau-level contribution (i.e., $k=0$) from the above equation, and we get

\begin{eqnarray}
    \label{eq30}
P_{\text{vac}}(S=1/2,B)= \frac{1}{2}\sum_{k=0}^{\infty}2\frac{|Q|B}{2\pi}
\int_{-\infty}^{\infty}\frac{dp_z}{2\pi} \nonumber \\
\bigg[E_{p,k}(B)-\frac{E_{p,0}(B)}{2}\bigg].
\end{eqnarray}
A dimensional regularization method~\cite{Peskin:1995} is used to regularize the ultraviolet divergence of vacuum pressure. In $d-\varepsilon$ dimension Eq.~(\ref{eq30}) can be written as

\begin{eqnarray}
\label{eq31}
P_{\text{vac}}(S=1/2,B)=\sum_{k=0}^{\infty}\frac{|Q|B}{2\pi}\mu^{\varepsilon}\int_{-\infty}^{\infty}\frac{d^{1-\varepsilon}p_z}{(2\pi)^{1-\varepsilon}} \nonumber\\ \bigg[\sqrt{p_z^2+m^2-2|Q|Bk}-\sqrt{p_z^2+m^2}\bigg],
\end{eqnarray}
 
In the preceding equation, $\mu$ sets the dimension to 1. The integration can be carried out using the usual $d-$dimensional formulas ~\cite{Peskin:1995, Ramond:2001}. 
\begin{equation}
\label{eq32}
\int_{-\infty}^{\infty}\frac{d^dp}{(2\pi)^d}\:\bigg[p^2+m^2\bigg]^{-A}=\frac{\Gamma[A-\frac{d}{2}]}{(4\pi)^{d/2}\Gamma[A](m^2)^{(A-\frac{d}{2})}}.
\end{equation}
 
Integration of the first term in Eq.~(\ref{eq31}) gives
\begin{eqnarray}
\label{eq33}
I_{1}=\sum_{k=0}^{\infty}\frac{|Q|B}{2\pi}\mu^{\varepsilon}\int_{-\infty}^{\infty}\frac{d^{1-\varepsilon}p_z}{(2\pi)^{1-\varepsilon}}\bigg[p_z^2+m^2-2|Q|Bk\bigg]^{\frac{1}{2}} \nonumber\\ 
=-\frac{(|Q|B)^2}{4\pi^2}\bigg(\frac{2|Q|B}{4\pi\mu}\bigg)^{-\frac{\varepsilon}{2}}\Gamma\bigg[-1+\frac{\varepsilon}{2}\bigg]\zeta\bigg[-1+\frac{\varepsilon}{2},x\bigg],
\end{eqnarray}

where we denote $x\equiv\frac{m^2}{2|Q|B}$. The Landau infinite sum has been illustrated in terms of the Riemann-Hurwitz $\zeta$ function
\begin{equation}
\label{eq34}
\zeta[z,x]=\sum_{k=0}^{\infty}\frac{1}{[x+k]^z},
\end{equation}
with the expansion~\cite{Digital:mathematica, Elizalde:1986},

\begin{equation}
\label{eq35}
\zeta\bigg[-1+\frac{\varepsilon}{2},x\bigg]\approx-\frac{1}{12}-\frac{x^2}{2}+\frac{x}{2}+\frac{\varepsilon}{2}\zeta^{'}(-1,x)+\mathcal{O}(\varepsilon^2)
\end{equation}
and the asymptotic behavior of the derivative~\cite{ Digital:mathematica, Elizalde:1986},

\begin{eqnarray}
\label{eq36}
\zeta'(-1,x) &=& \frac{1}{12} -\frac{x^2}{4} + \left( \frac{1}{12} -\frac{x}{2} + \frac{x^2}{2}\right) ln(x)+
\mathcal{O}(x^{-2}).\nonumber\\ 
\end{eqnarray}

The expansion of the $\Gamma$ function around some negative integers is given by
\begin{equation}
\label{eq37}
\Gamma\bigg[-1+\frac{\varepsilon}{2}\bigg]=-\frac{2}{\varepsilon}+\gamma-1+\mathcal{O}(\varepsilon),
\end{equation}
and,
\begin{equation}
\label{eq38}
\Gamma\bigg[-2+\frac{\varepsilon}{2}\bigg]=\frac{1}{\varepsilon}-\frac{\gamma}{2}+\frac{3}{4}+\mathcal{O}(\varepsilon).
\end{equation}

Here, $\gamma$ is the Euler constant. The limiting expression for natural is,
\begin{equation}
\label{eq39}
\lim_{\varepsilon\longrightarrow 0}a^{-\varepsilon/2}\approx 1-\frac{\varepsilon}{2}\text{ln}(a).
\end{equation} 

Equation~(\ref{eq33}) can be written as expressed as using the expansion of the $\Gamma$ function and $\zeta$ function: 
\begin{eqnarray}
\label{eq40}
I_1=-\frac{(|Q|B)^2}{4\pi^2}\bigg[-\frac{2}{\varepsilon}+\gamma-1+\text{ln}\bigg(\frac{2|Q|B}{4\pi\mu^2}\bigg)\bigg] \nonumber\\
\bigg[-\frac{1}{12}-\frac{x^2}{2}+\frac{x}{2}+\frac{\varepsilon}{2}\zeta^{'}(-1,x)+\mathcal{O}(\varepsilon^2)\bigg]
\end{eqnarray}
 
The second term in Eq.~(\ref{eq31}) can be simplified in the same way, and we obtain,
\begin{eqnarray}
\label{eq41}
I_{2}=\sum_{k=0}^{\infty}\frac{|Q|B}{2\pi}\mu^{\varepsilon}\int_{-\infty}^{\infty}\frac{d^{1-\varepsilon}p_z}{(2\pi)^{1-\varepsilon}}\bigg[p_z^2+m^2\bigg]^{\frac{1}{2}} \nonumber\\
=\frac{(|Q|B)^2}{4\pi^2}\bigg[-\frac{x}{\varepsilon}-\frac{(1-\gamma)}{2}x+\frac{x}{2}\text{ln}\bigg(\frac{2|Q|B}{4\pi\mu^2}\bigg)+\frac{x}{2}\text{ln}(x)\bigg].\nonumber \\
\end{eqnarray}

Hence, the vacuum pressure in the presence of an external magnetic field becomes
\begin{eqnarray}
\label{eq42}
P_{\text{vac}}(S=1/2,B)=\frac{(|Q|B)^2}{4\pi^2}\bigg[\zeta^{'}(-1,x)-\frac{2}{12\varepsilon}-\frac{(1-\gamma)}{12}\nonumber\\
-\frac{x^2}{\varepsilon}-\frac{(1-\gamma)}{2}x^2+\frac{x}{2}\text{ln}(x)\nonumber\\
+\frac{x^2}{2}\text{ln}\bigg(\frac{2|Q|B}{4\pi\mu^2}\bigg)+\frac{1}{12}\text{ln}\bigg(\frac{2|Q|B}{4\pi\mu^2}\bigg)\bigg].\nonumber\\
\end{eqnarray}
 
Divergence is still evident in the preceding expression. As a result, we add and deduct the $B=0$ contribution from it. To carry out the renormalization of the $B>0$ pressure, the $B=0$ contribution must be determined. The vacuum pressure in $d=3-\varepsilon$ dimensions at $B=0$ is given by 

\begin{eqnarray}
\label{eq43}
P_{\text{vac}}(S=1/2,B=0)= \mu^{\varepsilon}\int\frac{d^{3-\varepsilon}p}{(2\pi)^{3-\varepsilon}}\:(p^2+m^2)^{\frac{1}{2}}\nonumber\\
=\frac{(|Q|B)^2}{4\pi^2}\bigg(\frac{2|Q|B}{4\pi\mu^2}\bigg)^{-\frac{\varepsilon}{2}}\Gamma\bigg(-2+\frac{\varepsilon}{2}\bigg)x^{2-\frac{\varepsilon}{2}}. \nonumber\\
\end{eqnarray}
 
Above Eq.~(\ref{eq43}) can be further simplified by using $\Gamma$-function expansion from Eq.~(\ref{eq37}),
\begin{eqnarray}
\label{eq44}
P_{\text{vac}}(S=1/2,B=0)&=&-\frac{(|Q|B)^2}{4\pi^2}x^2\bigg[\frac{1}{\varepsilon}+\frac{3}{4}-\frac{\gamma}{2} \nonumber\\
&-&\frac{1}{2}\text{ln}\bigg(\frac{2|Q|B}{4\pi\mu^2}\bigg)-\frac{1}{2}\text{ln}(x)\bigg]
\end{eqnarray}

Now, we add and subtract Eq.~(\ref{eq44}) from (\ref{eq42}), we get the regularized pressure with the vacuum part, and the magnetic field-dependent part separated as
\begin{eqnarray}
\label{eq45}
P_{\text{vac}}(S=1/2,B)&=&P_{\text{vac}}(1/2,B=0)+\Delta P_{\text{vac}}(1/2,B), \nonumber\\
\end{eqnarray}
 
where,
\begin{eqnarray}
\label{eq46}
\Delta P_{\text{vac}}(S=1/2,B)&=&\frac{(|Q|B)^2}{4\pi^2}\bigg[-\frac{2}{12\varepsilon}+\frac{\gamma}{12}\nonumber\\
&+&\frac{1}{12}\text{ln}\bigg(\frac{m^2}{4\pi\mu^2}\bigg)+\frac{x}{2}\text{ln}(x)-\frac{x^2}{2}\text{ln}(x)\nonumber\\
&+&\frac{x^2}{4}-\frac{\text{ln}(x)+1}{12}+\zeta^{'}(-1,x)\bigg].
\end{eqnarray}
 
The field contribution given by the Eq.~(\ref{eq46}) is, however, divergent due to the existence of the magnetic field-dependent term $\frac{B^2}{\varepsilon}$~\cite{Schwinger:1951nm,Elmfors:1993bm,Andersen:2011ip}. We eliminate this divergence by redefining field-dependent pressure contribution to include magnetic field contribution.

\begin{equation}
\label{eq47}
\Delta P_{\text{vac}}^{r}=\Delta P_{\text{vac}}(B)-\frac{B^2}{2}.
\end{equation}
 
 The divergences are absorbed into the renormalization of the electric charge and the magnetic field strength ~\cite{Endrodi:2013cs},

\begin{equation}
\label{eq48}
B^2=Z_{e}B_r^2; \hspace{0.5cm} e^2=Z_e^{-1}e_r^2; \hspace{0.5cm} e_rB_r=|Q|B,
\end{equation}
 
 where the electric charge renormalization constant is
 
\begin{eqnarray}
\label{eq49}
Z_{e}\bigg(S=\frac{1}{2}\bigg)=1+\frac{1}{2}e_r^2\bigg[-\frac{2}{12\varepsilon}+\frac{\gamma}{12}+\frac{1}{12}\text{ln}\bigg(\frac{m_{*}}{4\pi\mu^2}\bigg)\bigg].\nonumber\\
\end{eqnarray}
 
We fix $m_*=m$, i.e. the particle's physical mass. Thus, the contribution of the renormalized field-dependent pressure in the absence of a pure magnetic field ($\frac{B^{2}}{2}$) is,
 
 \begin{eqnarray}
\label{eq50}
\Delta P_{\text{vac}}^r(S=1/2,B)&=&\frac{(|Q|B)^2}{4\pi^2}\bigg[\zeta^{'}(-1,x)+\frac{x}{2}\text{ln}(x)\nonumber\\
&-&\frac{x^2}{2}\text{ln}(x)+\frac{x^2}{4}-\frac{\text{ln}(x)+1}{12}\bigg]
\end{eqnarray}
 
Using a similar technique, the renormalized magnetic field-dependent pressure for spin-zero and spin-one particles can be calculated. These terms are crucial in determining the magnetization of hadronic matter. The vacuum pressure is affected by the charge, mass, and spin of the particles. As a result, the total vacuum pressure of a hadron gas is calculated by adding the vacuum pressures of all particles taken into account.

For spin-zero particles, the regularized vacuum pressure is
\begin{eqnarray}
\label{eq51}
\Delta P_{\text{vac}}^r(s=0,B)&=&-\frac{(|Q|B)^2}{8\pi^2}\bigg[\zeta^{'}(-1,x+1/2)-\frac{x^2}{2}\text{ln}(x)\nonumber\\
&+&\frac{x^2}{4}+\frac{\text{ln}(x)+1}{24}\bigg].
\end{eqnarray}

Similarly, for spin-one particles,
\begin{eqnarray}
\label{eq52}
\Delta P_{\text{vac}}^r(s=1,B)&=& -\frac{3}{8\pi^2}(|Q|B)^2\bigg[\zeta^{'}(-1,x-1/2)\nonumber\\
&+&\frac{(x+1/2)}{3}\text{ln}(x+1/2)\nonumber\\
&+&\frac{2}{3}(x-1/2)\text{ln}(x-1/2)-\frac{x^2}{2}\text{ln}(x)\nonumber\\
&+&\frac{x^2}{4}-\frac{7}{24}(\text{ln}(x)+1)\bigg].
\end{eqnarray}
 
 So, the total magnetic field-dependent vacuum pressure becomes
\begin{eqnarray}
\label{eq53}
\Delta P_{\text{vac}} &=& \Delta P_{\text{vac}}^r(s=0,B)+\Delta P_{\text{vac}}^r(S=1/2,B)\nonumber\\
&+&\Delta P_{\text{vac}}^r(s=1,B).\nonumber\\
\end{eqnarray}

After computing the total vacuum pressure, the system's vacuum magnetization can be computed as follows:
\begin{equation}
\label{eq54}
\Delta \mathcal{M}_{\text{vac}} = \frac{\partial (\Delta P_{\text{vac}})}{\partial (|Q|B)}.
\end{equation}

The explicit calculations of $\Delta \mathcal{M}_{\text{vac}}$ for spin-0, spin-1/2 and spin-1 particle are shown in Appendix ~\ref{appe4}.
By using the formalism mentioned in the above two sections, we estimate various thermodynamic observables for a hadron gas with van der Waals interaction.

\begin{figure*}[!]
\begin{subfigure}{0.35\textwidth}
    \includegraphics[scale=0.35]{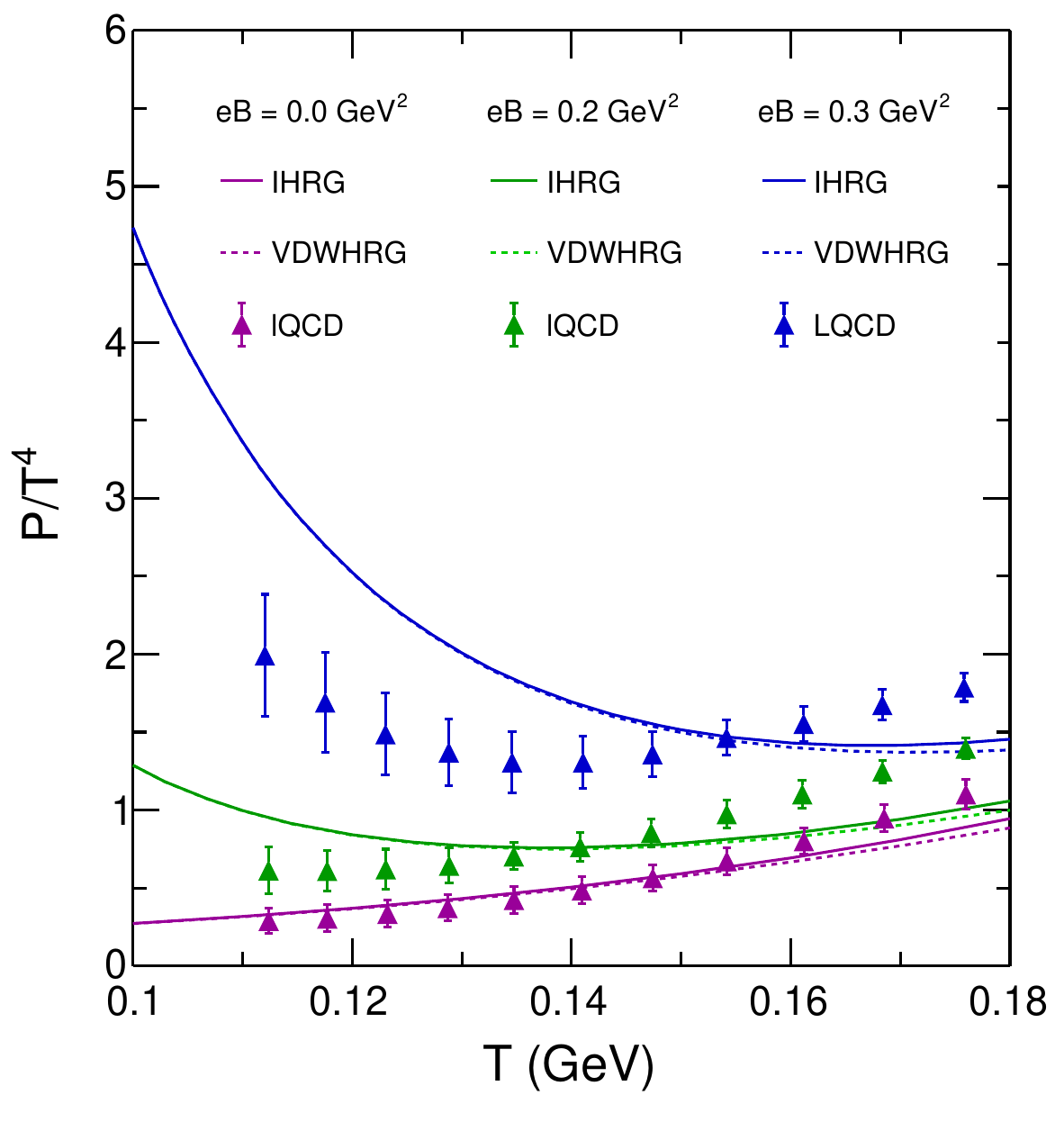}
    \caption{} \label{fig:1a}
  \end{subfigure}%
  \hspace*{1.5cm}   
\begin{subfigure}{0.35\textwidth}
\includegraphics[scale=0.35]{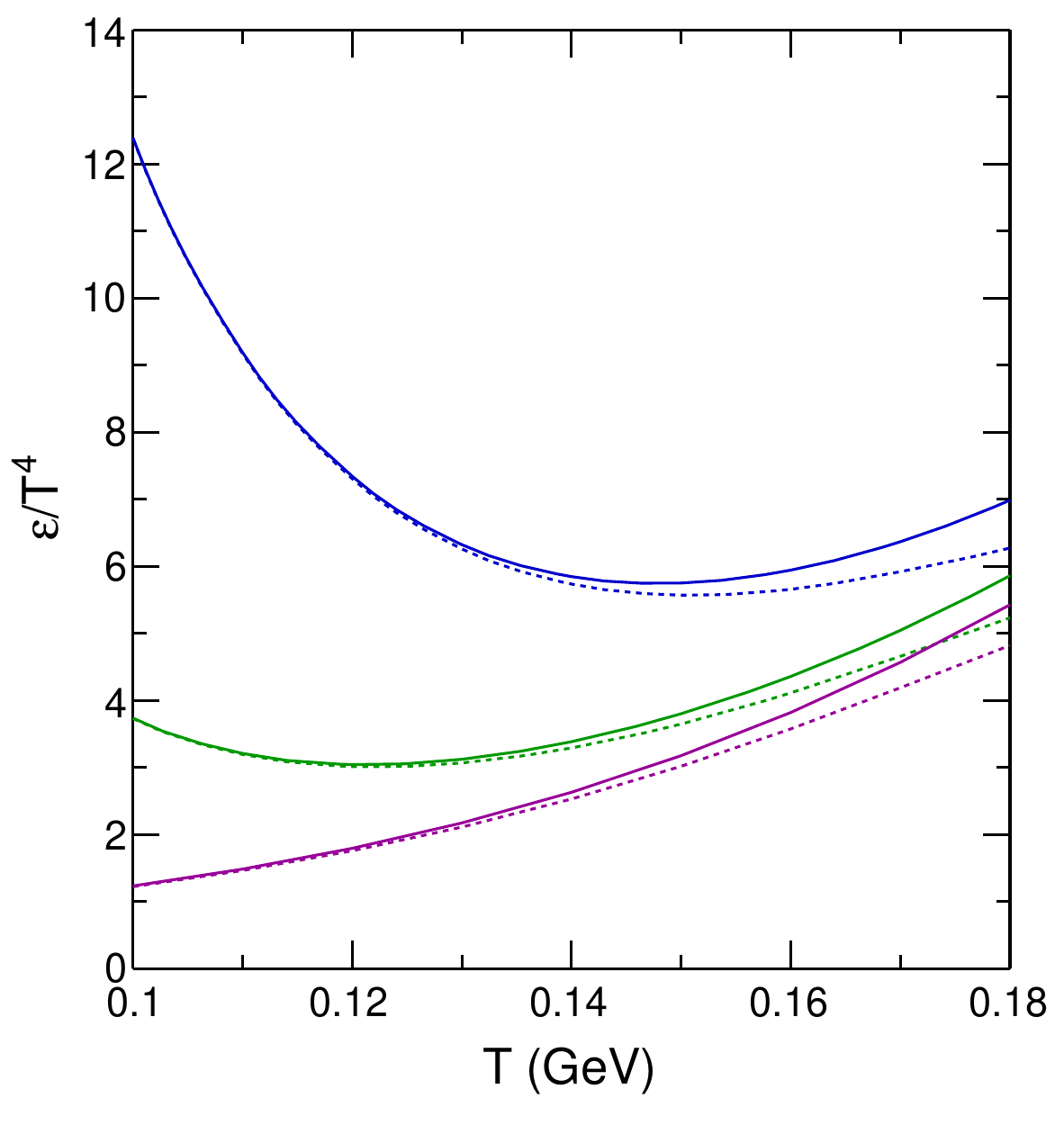}
 \caption{} \label{fig:1b}
  \end{subfigure}%
  
  \begin{subfigure}{0.35\textwidth}
\includegraphics[scale=0.35]{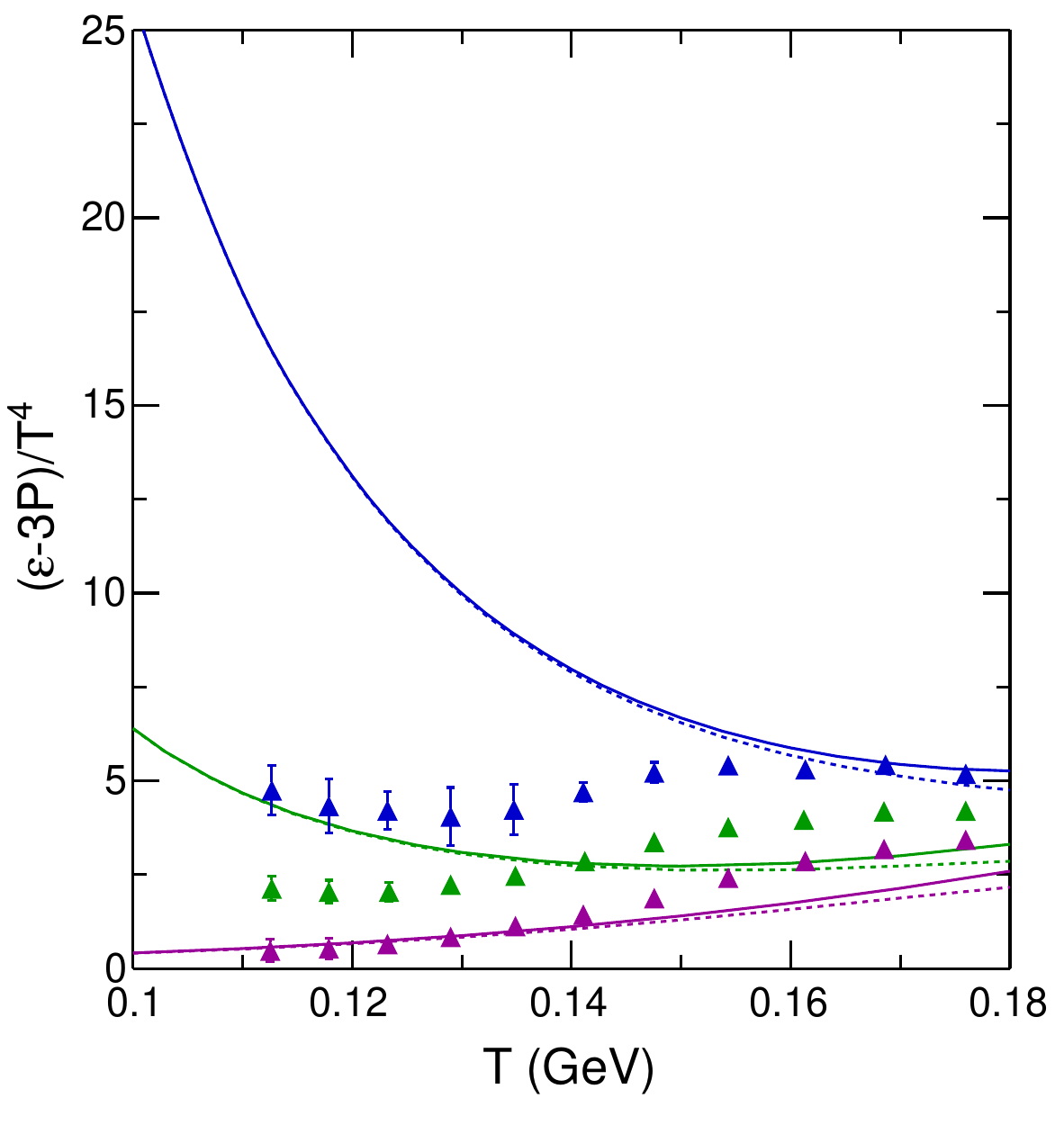}
\caption{} \label{fig:1c}
  \end{subfigure}%
  \hspace*{1.5cm}  
  \begin{subfigure}{0.35\textwidth}
 \includegraphics[scale=0.35]{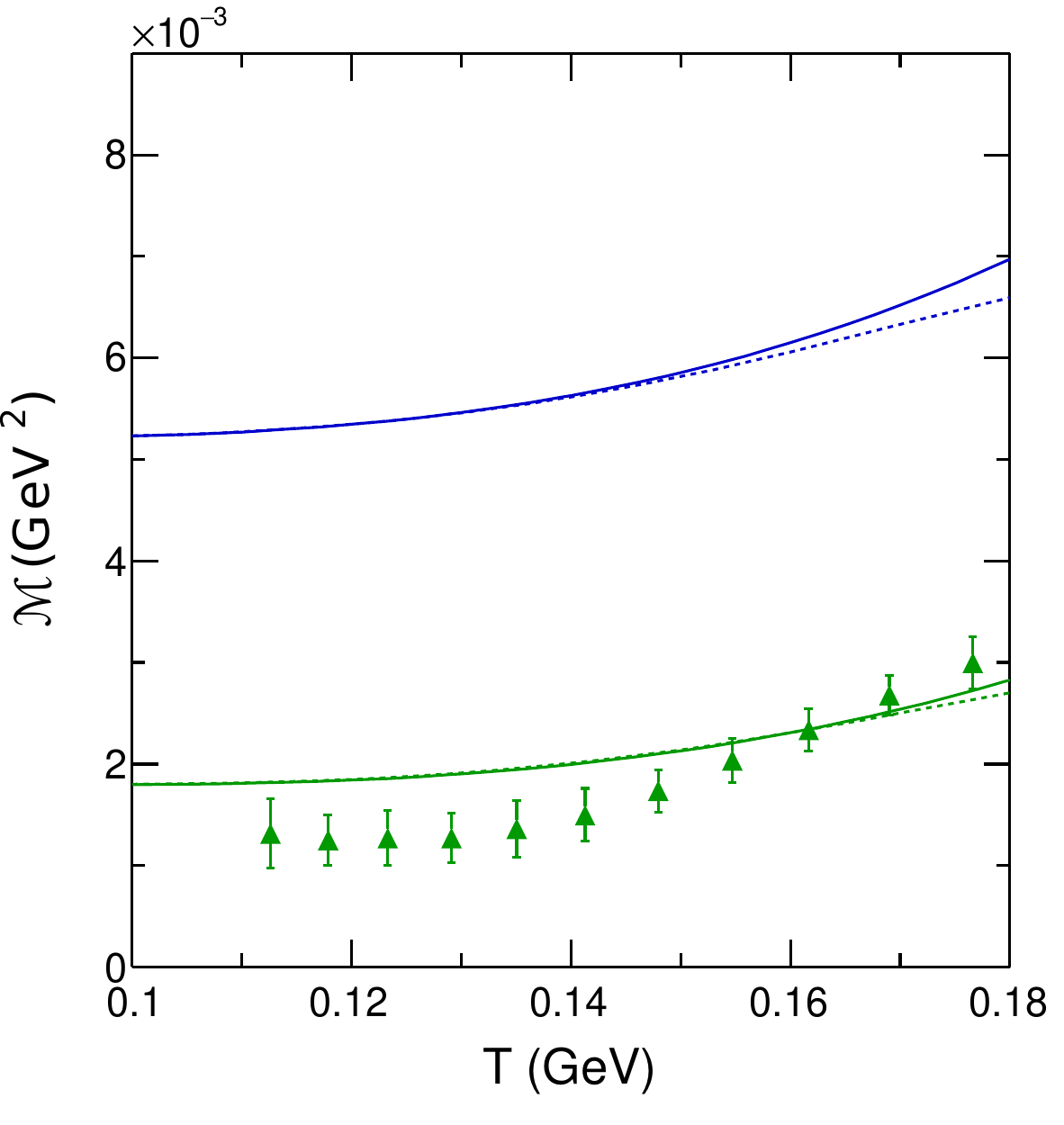}
 \caption{} \label{fig:1d}
  \end{subfigure}%

\begin{subfigure}{0.35\textwidth}
 \includegraphics[scale=0.35]{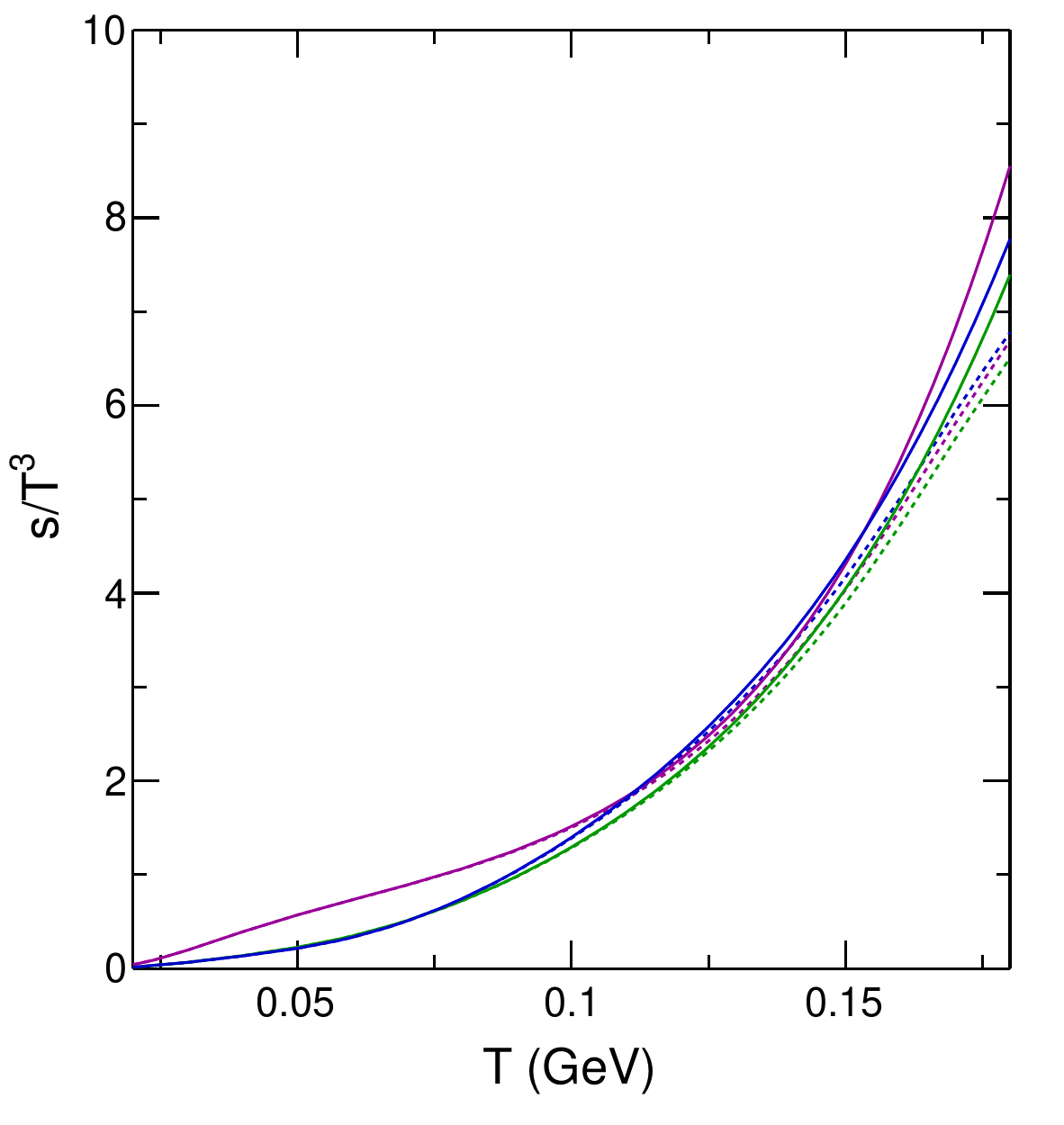}
\caption{} \label{fig:1e}
  \end{subfigure}%
\hspace*{1.5cm} 
\begin{subfigure}{0.35\textwidth}
   \includegraphics[scale=0.35]{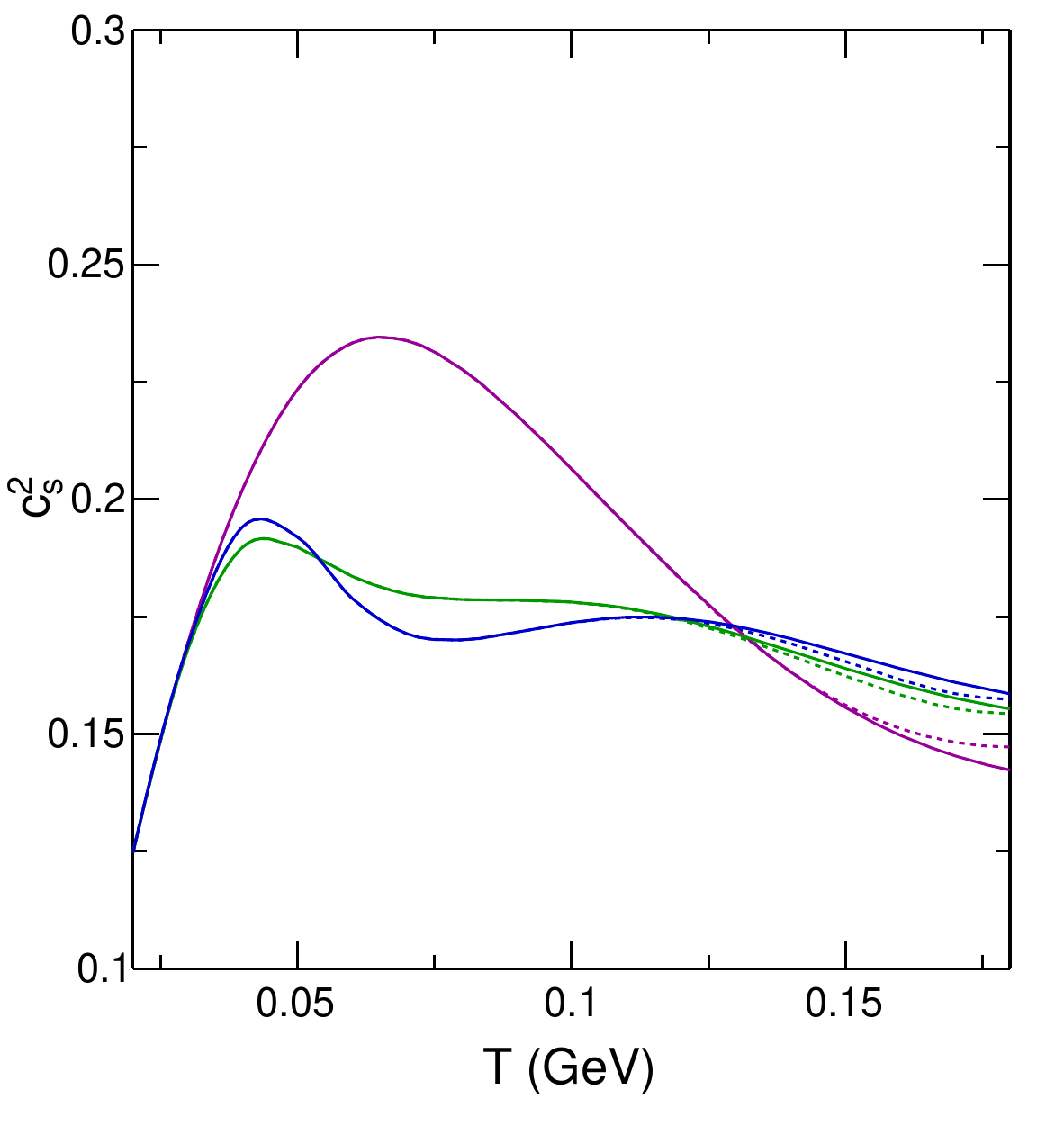}
\caption{} \label{fig:1f}
  \end{subfigure}%
   
  \caption{ The equation of state in the ideal HRG (VDWHRG) model is shown in a solid line (dotted line). The variation (from left to right and downwards) of normalized pressure, energy density, trace anomaly, magnetization, entropy density, and squared speed of sound as functions of temperature at zero baryochemical potential ($\mu_{B} $= 0 GeV ), for eB =0 $\rm GeV^{2} $ (magenta), eB= 0.2 $\rm GeV^{2} $(green), and eB = 0.3 $\rm GeV^{2} $(blue). The lattice data are taken from Ref.~\cite{Bali:2014kia}.}
  
  \label{fig:1}
\end{figure*}


\section{Results and Discussion}
\label{res}

In the present section, we discuss the results obtained from this study. It is important to note that we obtain all the results at $\mu_{Q} = 0$ and $\mu_{S} =0 $. So the chemical potential of the system is only due to $\mu_{B}$. We explore the effect of the magnetic field on thermodynamic observables at both zero and finite baryon chemical potential values. This study includes all hadrons and resonances of spin-0, spin-1/2, and spin-1 up to a mass cutoff of 2.25 GeV according to Particle Data Group \cite{PDG2016}. For any nonzero magnetic field, the spin-3/2 resonances give a negative contribution to the pressure, indicating an instability in the theory. This instability suggests that Eq.~(\ref{eq10}) which describes the dispersion relation in the HRG model is not applicable for spin-3/2 resonances. This is discussed in detail in Ref.~\cite{Endrodi:2013cs}. For the above-mentioned reason, we do not consider resonances with spin-3/2 or higher in the present model. One can obtain the van der Waals parameters by fitting the thermodynamic quantities, such as energy density, pressure, etc., in the VDWHRG model to the available lattice QCD data at zero magnetic fields \cite{Sarkar:2018mbk}. In principle, the van der Waals parameters should change in the presence of the magnetic field as well as the baryochemical potential. However, changing $a$ and $b$ parameters as a function of $eB$ as well as $\mu_{B}$ is nontrivial. We have neglected such dependencies in the current study. We calculate the thermodynamic quantities such as pressure, energy density, entropy density, specific heat, and squared speed of sound using their corresponding formulas as given in Sec.\ref{formulation} at zero and finite magnetic fields in the ideal HRG and VDWHRG models.

 In the present work, we examine two different values of magnetic fields, i.e., $eB = 0.2 $ GeV$^{2}$ and $eB = 0.3 $ GeV$^{2}$ for our study. In the presence of a finite magnetic field, the system's total pressure contains a contribution from both the vacuum and the thermal part, while there is no such vacuum-pressure contribution for a vanishing magnetic field. So, at  B$\neq$ 0 and T=0, the system has some nonvanishing pressure called vacuum pressure~\cite{Endrodi:2013cs, Kadam:2019rzo}. The vacuum pressure for spin-0, spin-1/2, and spin-1 particles is calculated using Eqs.~(\ref{eq50}),~(\ref{eq51}), and~(\ref{eq52}). It is found that the vacuum pressure is positive for spin-0, spin-1/2, and spin-1 particles. The total vacuum pressure is obtained by summing over all spin states.
 \begin{figure*}[!]
\centering
\begin{subfigure}{0.35\textwidth}
     \includegraphics[scale=0.35]{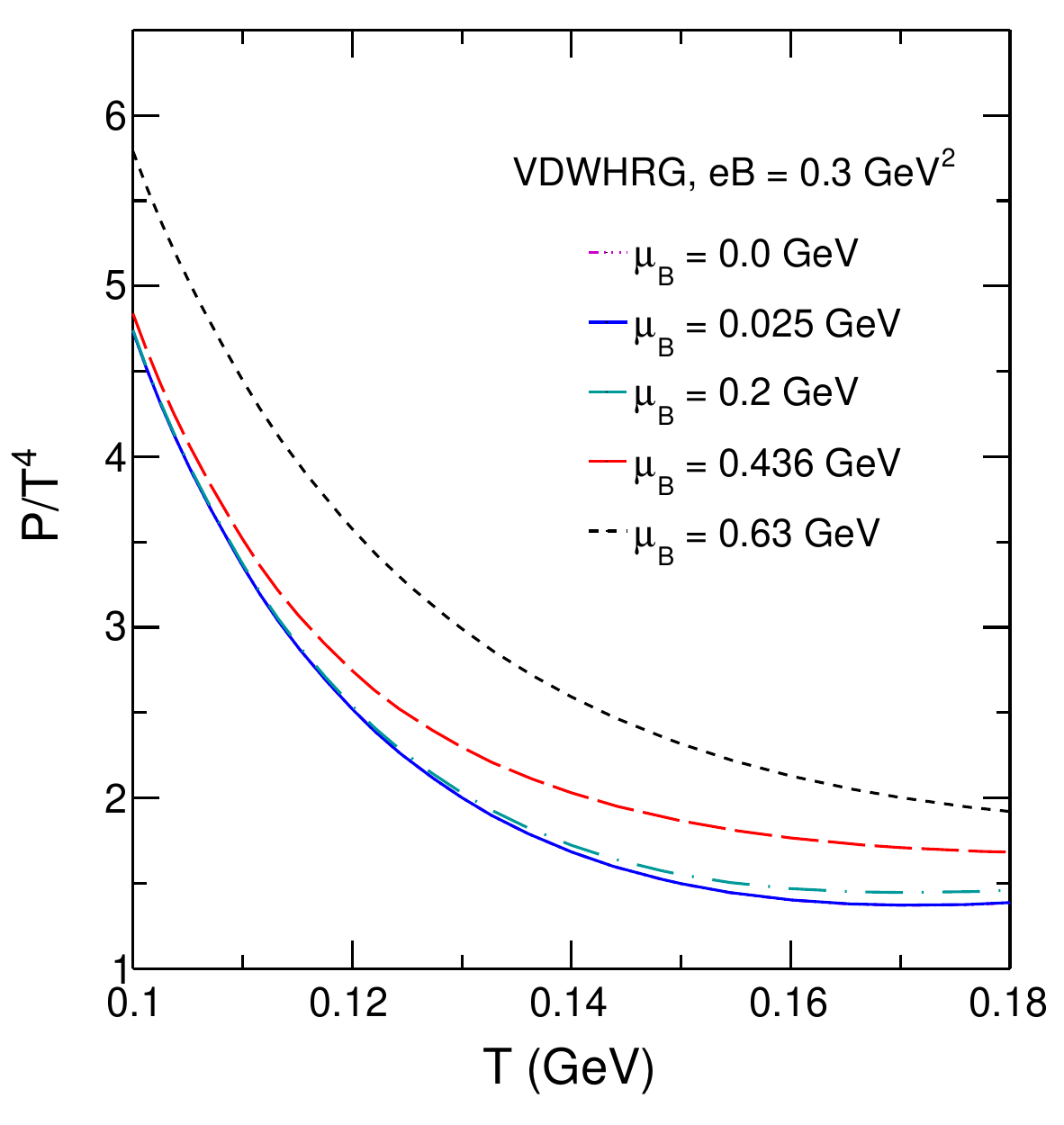}
    \caption{} \label{fig:1a}
  \end{subfigure}%
  \hspace*{1.5cm}   
\begin{subfigure}{0.35\textwidth}
\includegraphics[scale=0.35]{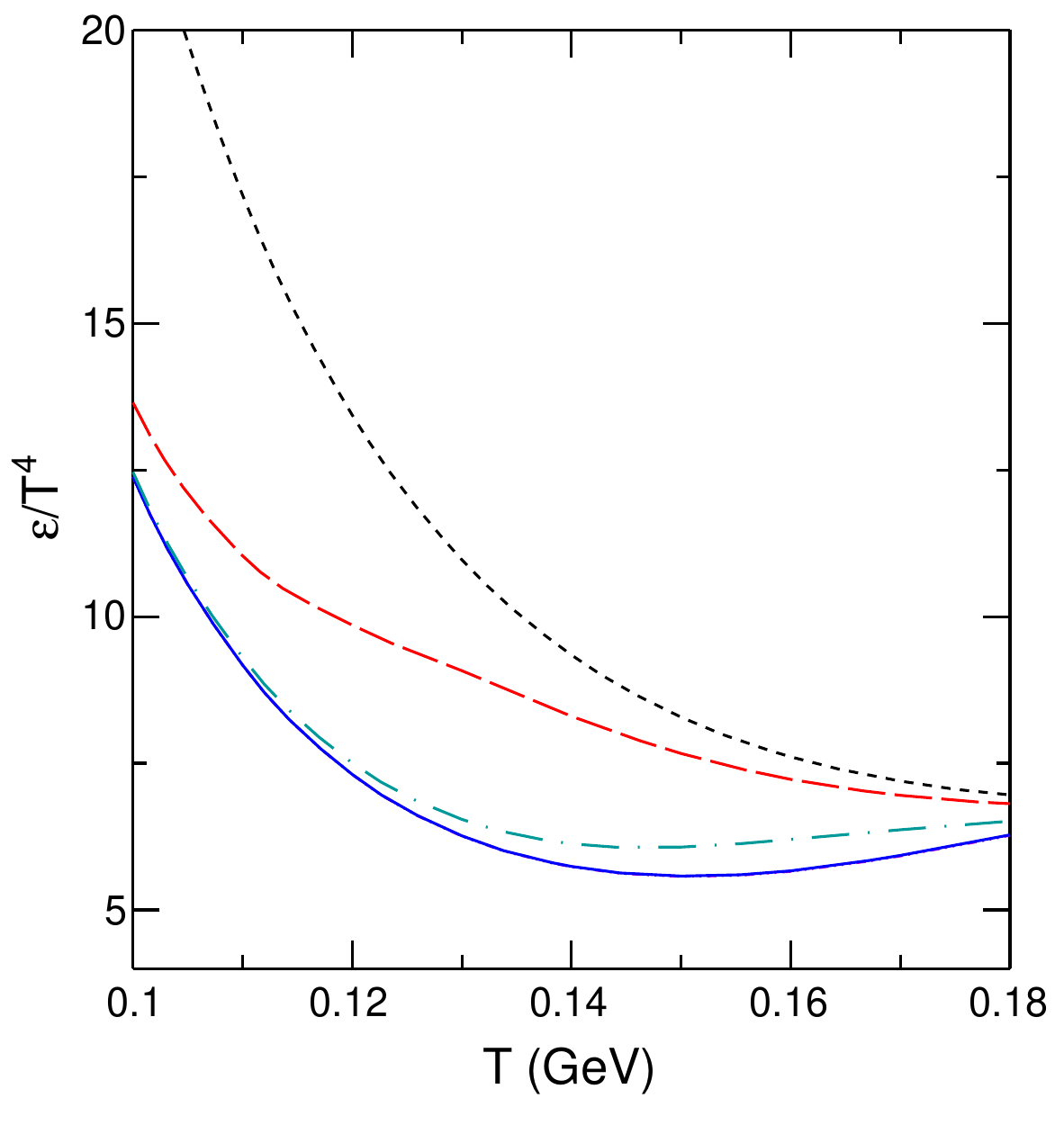}
 \caption{} \label{fig:1b}
  \end{subfigure}%

  \begin{subfigure}{0.35\textwidth}
    \includegraphics[scale=0.35]{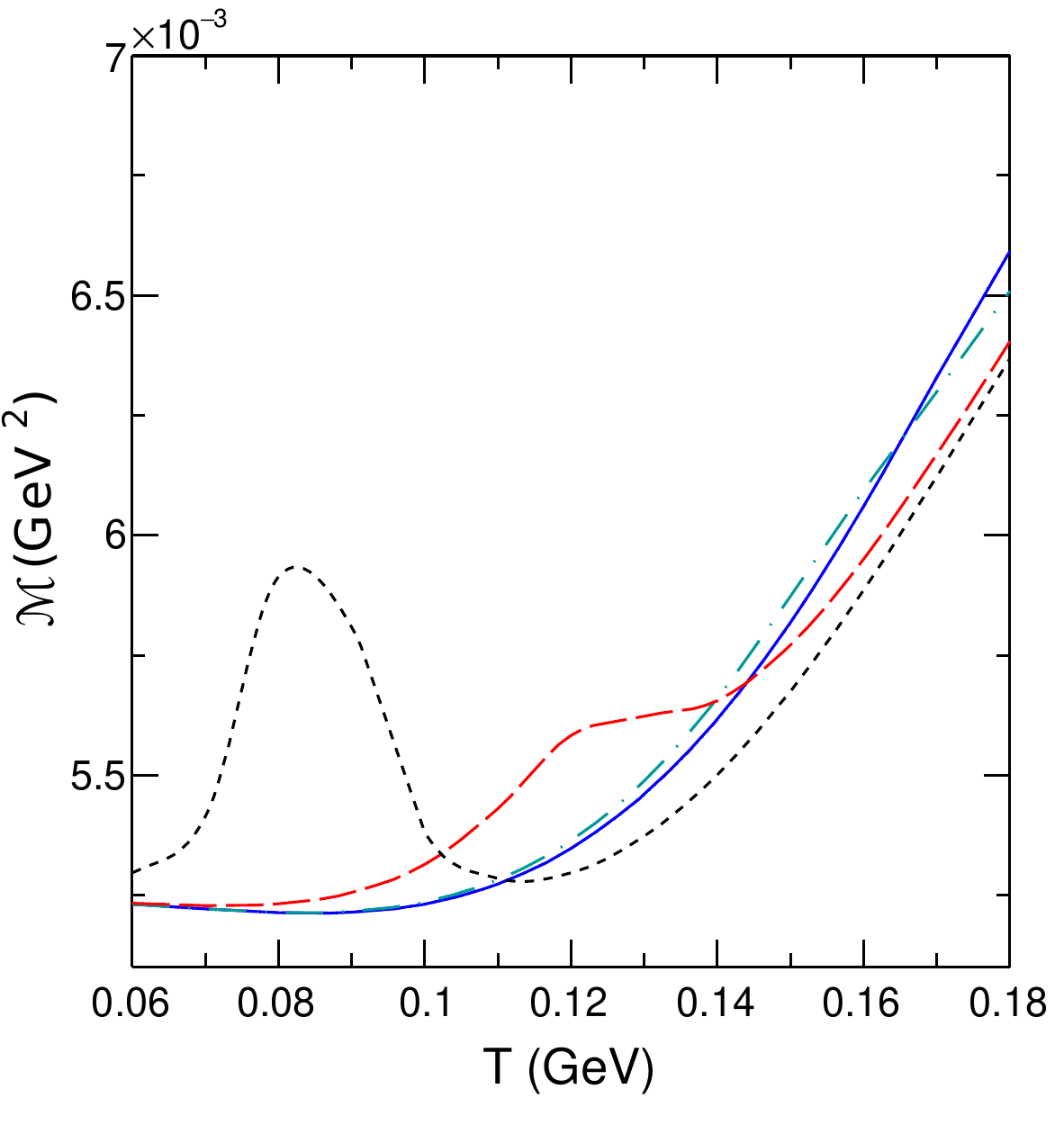}
    \caption{} \label{fig:1c}
  \end{subfigure}%
  \hspace*{1.5cm}   
\begin{subfigure}{0.35\textwidth}
 \includegraphics[scale=0.35]{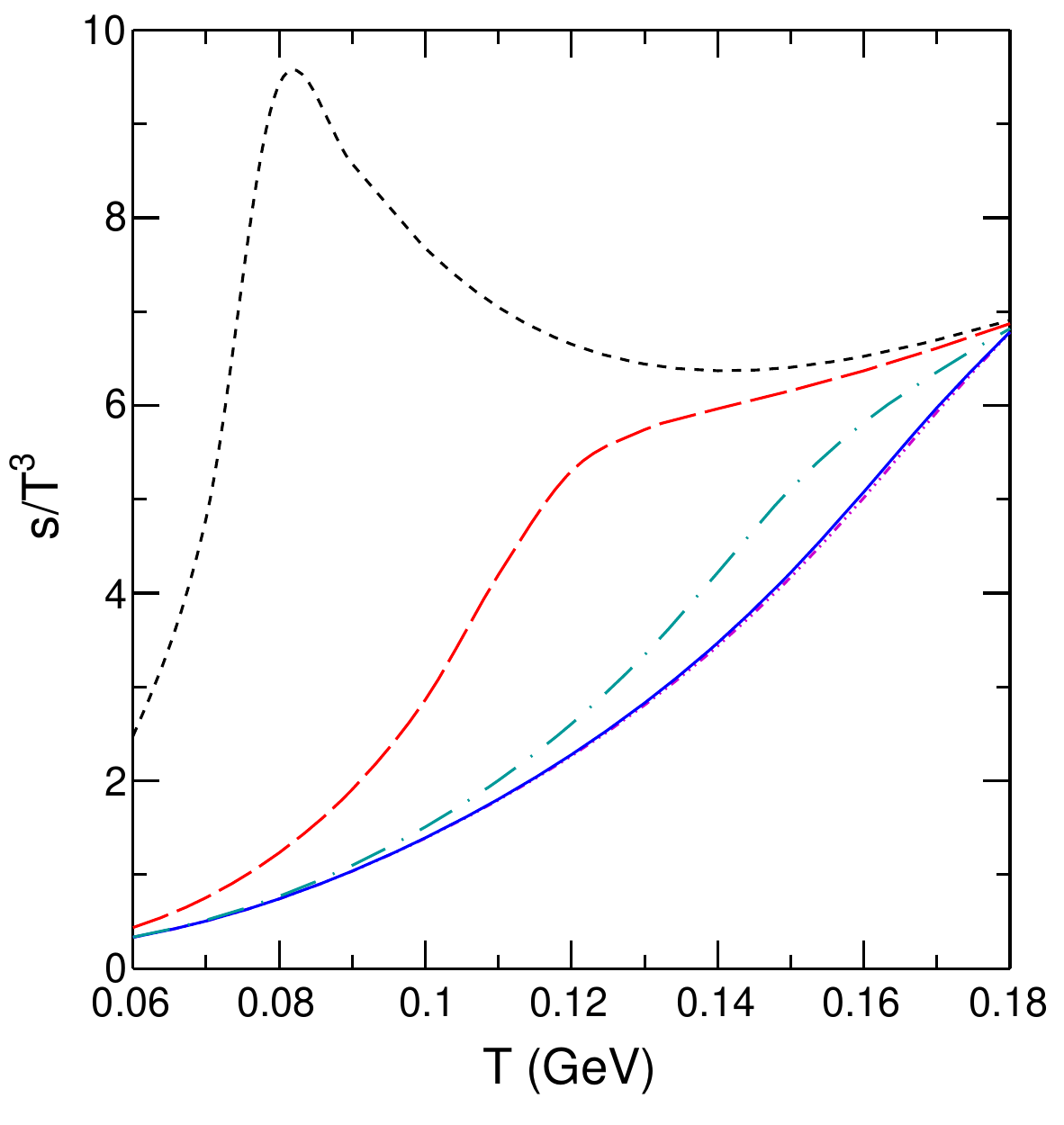}
 \caption{} \label{fig:1d}
  \end{subfigure}%
 
  \begin{subfigure}{0.35\textwidth}
    \includegraphics[scale=0.35]{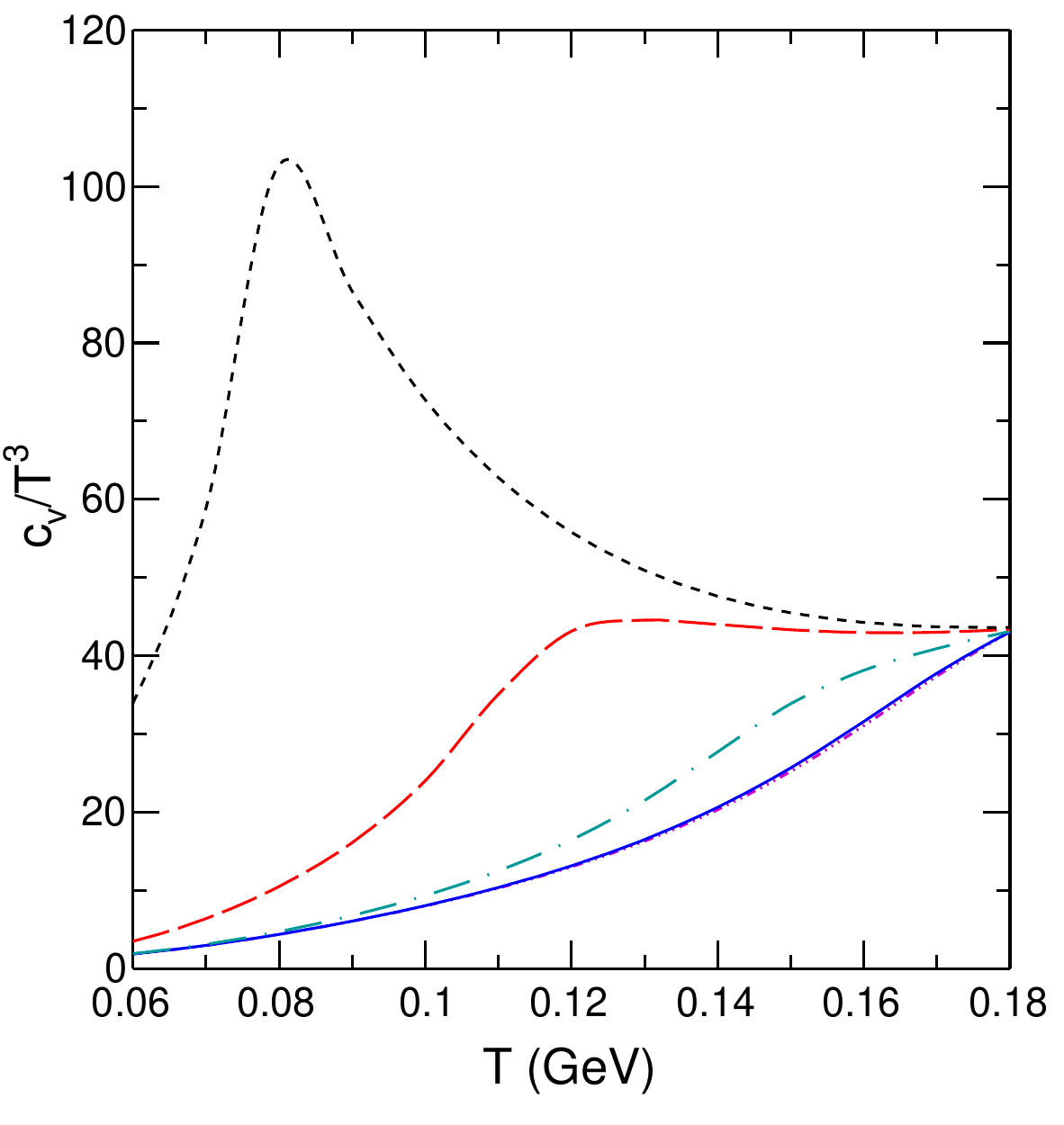}
    \caption{} \label{fig:1e}
  \end{subfigure}%
  \hspace*{1.5cm}   
\begin{subfigure}{0.35\textwidth}
  \includegraphics[scale=0.35]{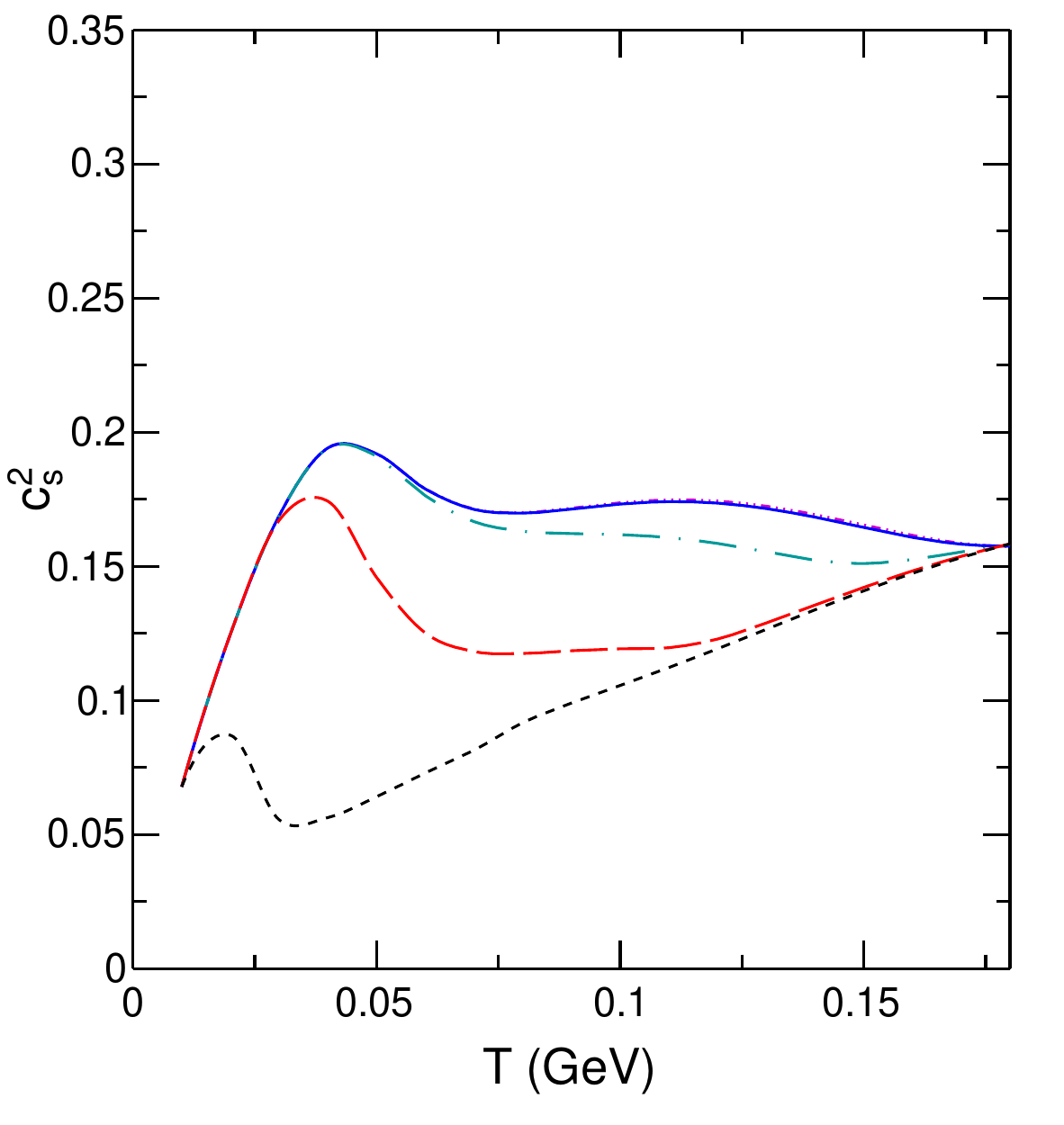}
 \caption{} \label{fig:1f}
  \end{subfigure}%
 
 \caption{The variation (from left to right and downwards) of normalized pressure, energy density, magnetization, entropy density, specific heat, and squared speed of sound as functions of temperature at different baryochemical potentials for eB = 0.3 $\rm GeV^{2} $.}
  \label{fig:2}

\end{figure*}
In Fig.\ref{fig:1}(a), we show the scaled pressure as a function of temperature in ideal HRG and VDWHRG for both zero and finite magnetic fields and compare it with the lQCD data. We observe that the pressure calculated in HRG and VDWHRG slightly deviates from the lQCD calculation, but the temperature dependence seems to be preserved. This deviation at high temperatures may be due to the fact that we are not considering higher-spin states in our calculations. One can observe that the normalized pressure increases with the temperature almost monotonically for a zero magnetic field, while for a finite magnetic field, it diverges at a lower temperature due to a finite vacuum contribution to the total pressure, both for the HRG and VDWHRG models. The pressure in the VDWHRG model is found to be suppressed slightly compared to the HRG model. However, we found that the total pressure of the system (without scaling with $T^{4}$) increases with temperature with an increase in a magnetic field. The lightest spin-0 particles [mainly dominated by pions($\pi^{\pm}, \pi^{0}$)] have more contribution towards pressure compared to heavier spin-1 ($\rho^{\pm},\rho^{0}$) and spin-1/2 [proton(p), neutron(n)] particles. In addition, it is noteworthy to mention that at lower temperatures, the thermal part of the pressure in the presence of a magnetic field is smaller than the pressure at a zero magnetic field. The vacuum pressure increases with an increase in the magnetic field, which is responsible for the monotonic increase in pressure with the magnetic field.
 
 \begin{figure*}[ht!]
\centering
\begin{subfigure}{0.35\textwidth}
    \includegraphics[scale=0.35]{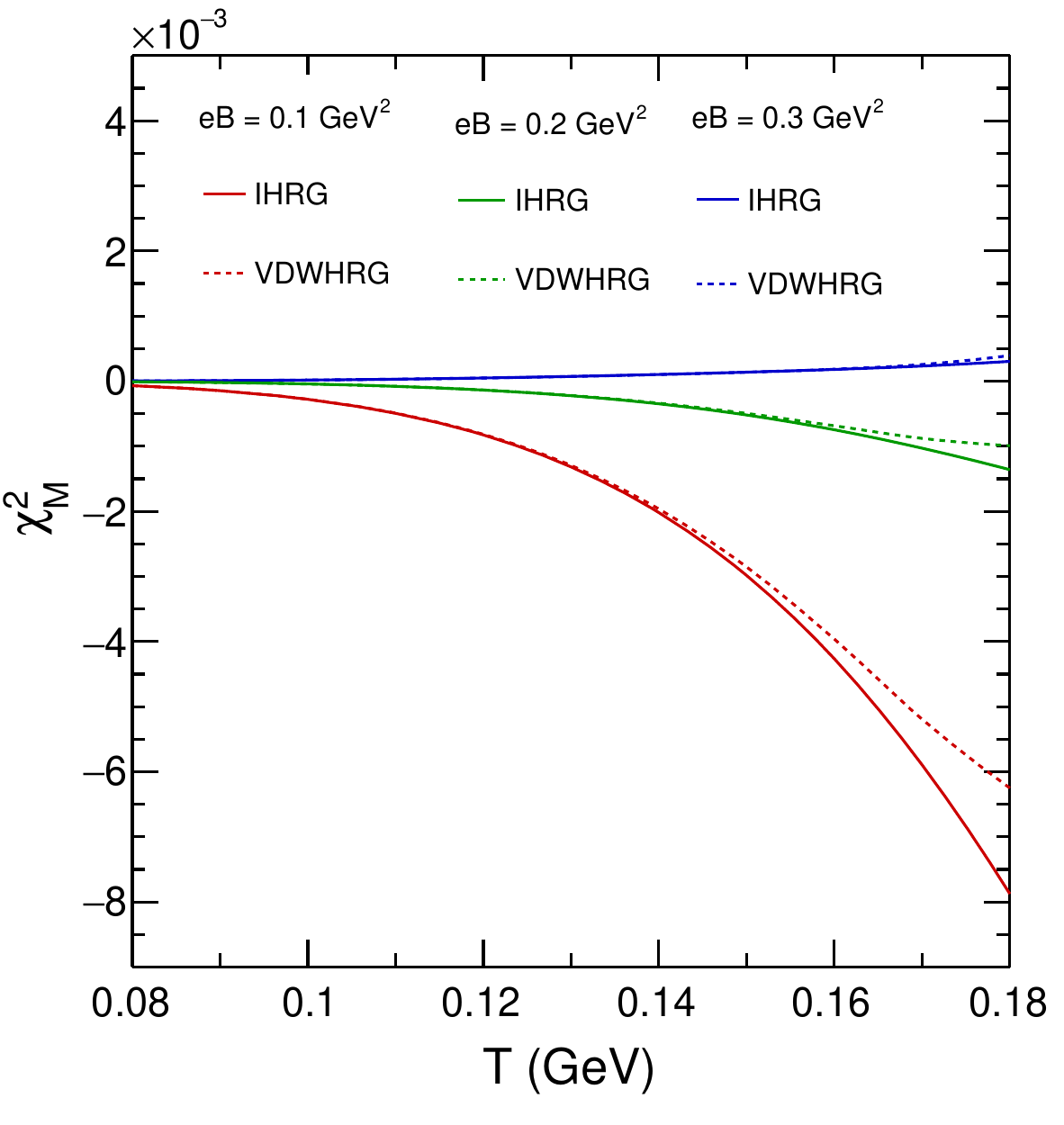}
    \caption{} \label{fig:4a}
  \end{subfigure}%
  \hspace*{1.5cm}   
\begin{subfigure}{0.35\textwidth}
\includegraphics[scale=0.35]{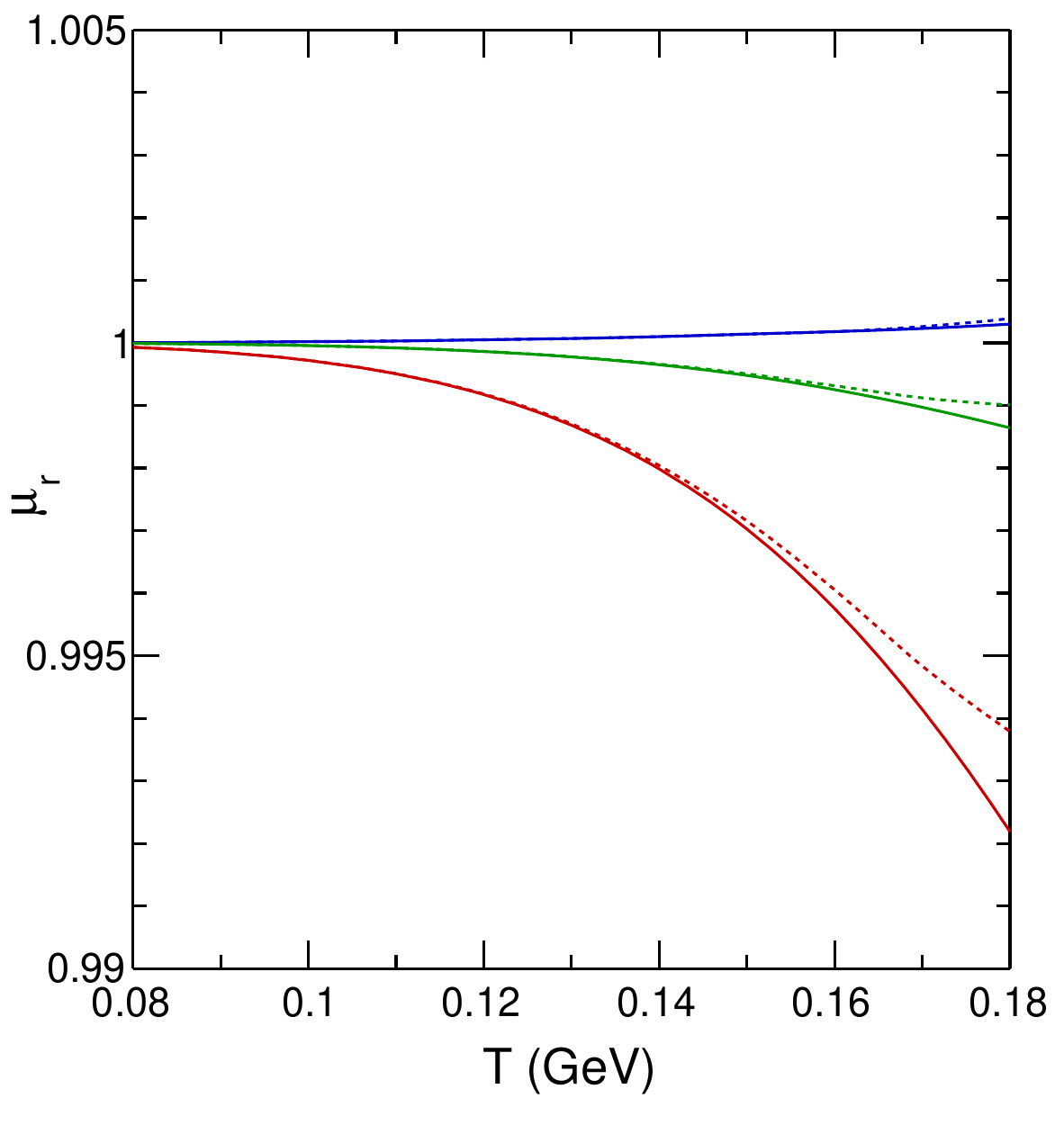} 
 \caption{} \label{fig:4b}
  \end{subfigure}%
\caption{ Magnetic susceptibility (left panel) and magnetic permeability (right panel) as functions of temperature for $eB$ = 0.1, 0.2, and 0.3 $\rm GeV^{2} $.}
  \label{fig:3}
\end{figure*}

\begin{figure}[ht!]

    \includegraphics[scale=0.35]{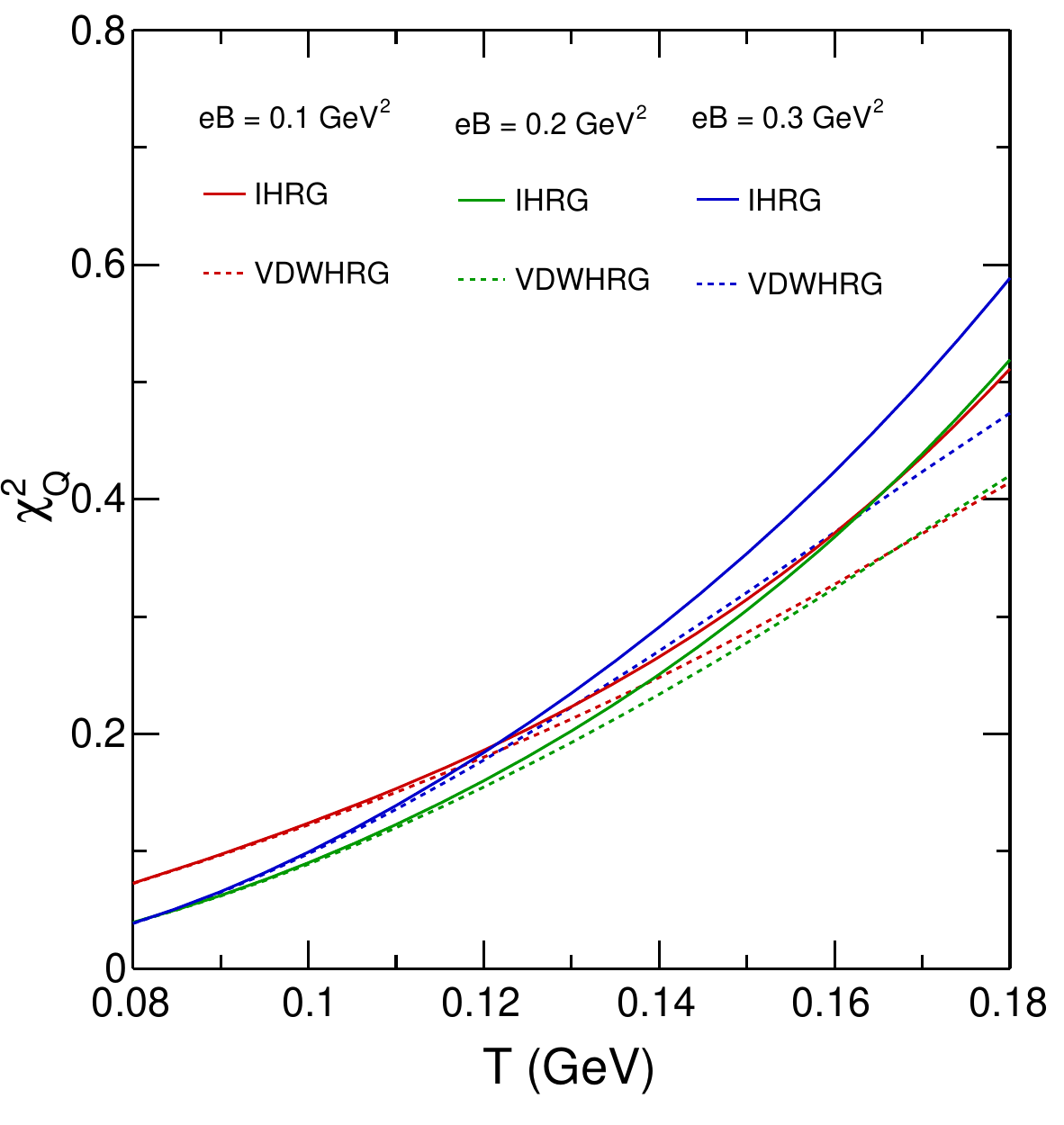}

  
 \caption{Electrical susceptibility as a function of temperature for $eB$ = 0.1, 0.2, 0.3 $\rm GeV^{2} $.}
  \label{fig:4}
\end{figure}

The total energy density in the presence of a magnetic field takes the form $ \varepsilon^{total} = \varepsilon + QB.\mathcal{M} $ \cite{Tawfik:2016lih}, where $\varepsilon^{total}$ and $\varepsilon$  represent the total energy density in the presence and absence of a magnetic field, respectively. 
 Figure~\ref{fig:1}(b) illustrates the variation of $\varepsilon/T^{4}$ as a function of $T$ along with the magnetic field in the ideal HRG and VDWHRG models. The $\varepsilon/T^{4}$ is found to increase with magnetic field for a fixed value of temperature. The $\varepsilon/T^{4}$  also exhibits divergence behavior at lower $T$ similar to that of $P/T^{4}$. It is found that there is a significant contribution of interactions after $T$ = 130 GeV, as shown in Fig.~\ref{fig:1}(b). The energy density is found to be suppressed at higher temperatures in the VDWHRG model due to dominating repulsive interactions.

The variation of the interaction measure (or normalized trace anomaly) as a function of temperature in the presence of a magnetic field is shown in Fig.~\ref{fig:1}(c). It can be directly derived from the energy-momentum tensor $T_{\mu}^{\nu}$), and it is sensitive to the massive hadronic states~\cite{Tawfik:2013eua}. For a perfect fluid, it is the sum of all diagonal elements of $T_{\mu}^{\nu}$. This parameter helps to determine the degrees of freedom of the system. We observe that the normalized trace anomaly diverges at a very low temperature, similar to the pressure and energy density. The magnetic field dependence of normalized trace anomaly is similar to normalized pressure and energy density and is comparable with the lQCD data~\cite{Bali:2014kia}.

 Figure~\ref{fig:1}(d) depicts the variation of magnetization as a function of temperature at zero $\mu_{B}$. The sign of magnetization defines the magnetic property of the system under consideration. A positive value of magnetization indicates the paramagnetic behavior of hadronic matter, which indicates the attraction of hadronic matter in an external magnetic field. This paramagnetic behavior of hadronic matter is observed in both ideal HRG and VDWHRG models. From Fig.~\ref{fig:1}(d), it is observed that magnetization has a monotonic behavior with increasing temperature. This magnetization contains contributions from both thermal and vacuum parts. The vacuum part of magnetization is calculated using Eq. (\ref{eq54}). The magnetization obtained for eB =0.2 $\rm GeV^{2} $ in HRG and VDWHRG model reasonably agrees with that of the lQCD simulation. At very low temperatures, the thermal part of magnetization is significantly less because of the lower abundance of charged hadrons. In addition to that, it is also important to note that the magnetization of charged pseudoscalar mesons (spin-0) is found to be negative. The magnetization of the hadronic matter becomes positive when the vector mesons (spin-1) and spin-1/2  baryons populate the hadronic matter at higher temperatures. It is also noteworthy to point out that we found even though the thermal part of magnetization is negative at lower temperatures, the total value of magnetization is always positive due to the vacuum contribution. 
 
 Figure~\ref{fig:1}(e) shows the change in entropy density as a function of temperature at zero and a finite magnetic field. Entropy being the first derivative of pressure with respect to temperature, there is no vacuum contribution term in entropy density. The value of entropy density is very small (almost vanishes) at lower temperatures, and it starts to increase with temperature. One can also notice that entropy density shows minimal deviation with magnetic field even at high temperatures. The entropy density is found to be suppressed because of the magnetic field. The effect of interactions also suppresses the value of entropy density. This observation may be interesting for HIC experiments since entropy acts as a proxy for particle multiplicity. 
 Although there is no significant dependence of the magnetic field on entropy density, the effect of the magnetic field on the squared speed of sound can be clearly visualized from Fig. ~\ref{fig:1}(f). The variation of the squared speed of sound as a function of temperature and the magnetic field is depicted at $\mu_{B} = 0$, and we notice that the $c_{s}^{2}$ exhibits a dip towards lower temperatures with a magnetic field. The minimum position of $c_{s}^{2}$ indicates the deconfinement transition temperature $T_{c}$.

Furthermore, we explore the variation of thermodynamic quantities in the presence of a finite chemical potential and a finite magnetic field. Fig.~\ref{fig:2} depicts the variation of $P/T^{4}$, $\varepsilon/T^{4}$, $\mathcal{M}$, $s$, $c_{v}$ and $c_{s}^{2}$, respectively as functions of temperature for both finite values of chemical potential and magnetic field in the VDWHRG model. We set different values of $\mu_{B}$, starting from 0.025 to 0.63 GeV, which correspond to the LHC, RHIC, FAIR, and NICA experiments \cite{Tawfik:2016sqd, Braun:2001sqd, Cleymans:2006sqd, Khuntia:2019sqd} at external magnetic field $eB$ = 0.3 $\rm GeV^{2} $. It should be noted that the strength of eB also decreases with a decrease in collision energy. Here, we have not considered the variation of eB with collision energy, as it is not straightforward. One can observe that for lower values of chemical potential (up to 0.2 GeV), the behavior of thermodynamic quantities in the VDWHRG model is almost like that of the zero chemical potential, with a slight variation in magnitude. But, there is a change in the behavior of some thermodynamic quantities observed for the higher value of chemical potential with magnetic field $eB$ = 0.3 $\rm GeV^{2} $. From Fig. ~\ref{fig:2}(a), it is observed that $P/T^{4}$ decreases monotonically with temperature for different values of chemical potential at $eB$ = 0.3 $\rm GeV^{2}$. A similar kind of observation is made in the energy density, with a slight variation in its trend. The magnetization, entropy density, and specific heat are found to increase with increasing temperature for lower values of chemical potential, as shown in Figs.~\ref{fig:2}(b),(c),(d),(e), respectively. But for higher values of chemical potential, the trend seems very interesting. The monotonic decreasing (increasing) behavior starts deviating for chemical potential around 0.436 GeV and above, as depicted in energy density, magnetization, entropy density, and specific-heat plots. Magnetization and entropy density, being the first-order derivatives of pressure with respect to magnetic field and temperature, respectively, show the behavior approaching a first-order phase transition at the higher chemical potential. The dependence of chemical potential on the squared speed of sound is quite interesting, as shown in Fig. ~\ref{fig:2}(f). The squared speed of sound decreases with an increase in chemical potential, showing a minimum. This minimum position shifts towards lower temperatures with higher values of chemical potential.

In addition to thermodynamic results, it is crucial to understand the susceptibility of the medium under consideration, which is a sensitive probe for QCD phase transition. 
The magnetic susceptibility provides knowledge about the strength of the hadronic matter's induced magnetization. Its sign distinguishes diamagnet ($\chi_{\rm M}^{2}  < 0$), which expels the external field, from paramagnet ($\chi_{\rm M}^{2} > 0$), which favors energetic exposure to the background field. In literature, the magnetic susceptibility of the HRG model is calculated through different approaches~\cite{Bali:2020bcn, Bali:2014kia}. Magnetic field dependence of magnetic susceptibility is also reported in the PNJL model~\cite{Chaudhuri:2022oru}. Figure~\ref{fig:3}(a) shows the magnetic field dependence of magnetic susceptibility with temperature. Since many of the thermodynamic quantities, including the fluctuation of conserved charges, are unaffected by the vacuum part ~\cite{Bhattacharyya:2015pra}, we neglect the vacuum contribution of susceptibilities in this study. One can observe that the magnetic susceptibility is negative for a lower value of magnetic field (e.g., eB = 0.1 $\rm GeV^{2} $ and eB =0.2 $\rm GeV^{2} $), and its value tends towards positive for a higher magnetic field (eB = 0.3 $\rm GeV^{2} $) both for the ideal HRG and VDWHRG models. So a clear observation of the diamagnetic to paramagnetic transition happens in the VDWHRG model. It is quite an exciting consequence of the study of magnetic field dependence on magnetic susceptibility.

Taking magnetic susceptibility into account, one can calculate the magnetic permeability of the medium. The relative magnetic permeability is defined as $\mu_{r}$= $\frac{\mu}{\mu_{0}} = \frac{1}{1-e^2\chi_{M}^{2}}$~\cite{Bonati:2013lca,Bali:2014kia}. This combination is equivalent to the ratio of the magnetic induction with the external field~\cite{Bonati:2013lca,Bali:2014kia}.
Figure~\ref{fig:3}(b) shows the magnetic field dependence of relative magnetic permeability with temperature in the ideal HRG and VDWHRG models. It is observed that the relative magnetic permeability is close to unity at lower temperatures, and it starts deviating from unity (although the deviation is very small in magnitude) while going toward higher temperatures. The $\mu_{r}$ decreases with an increase in temperature at the lower magnetic field. Further, it starts to increase with the rise in the magnetic field.

\begin{figure*}[ht!]
\centering
\begin{subfigure}{0.35\textwidth}
   \includegraphics[scale=0.35]{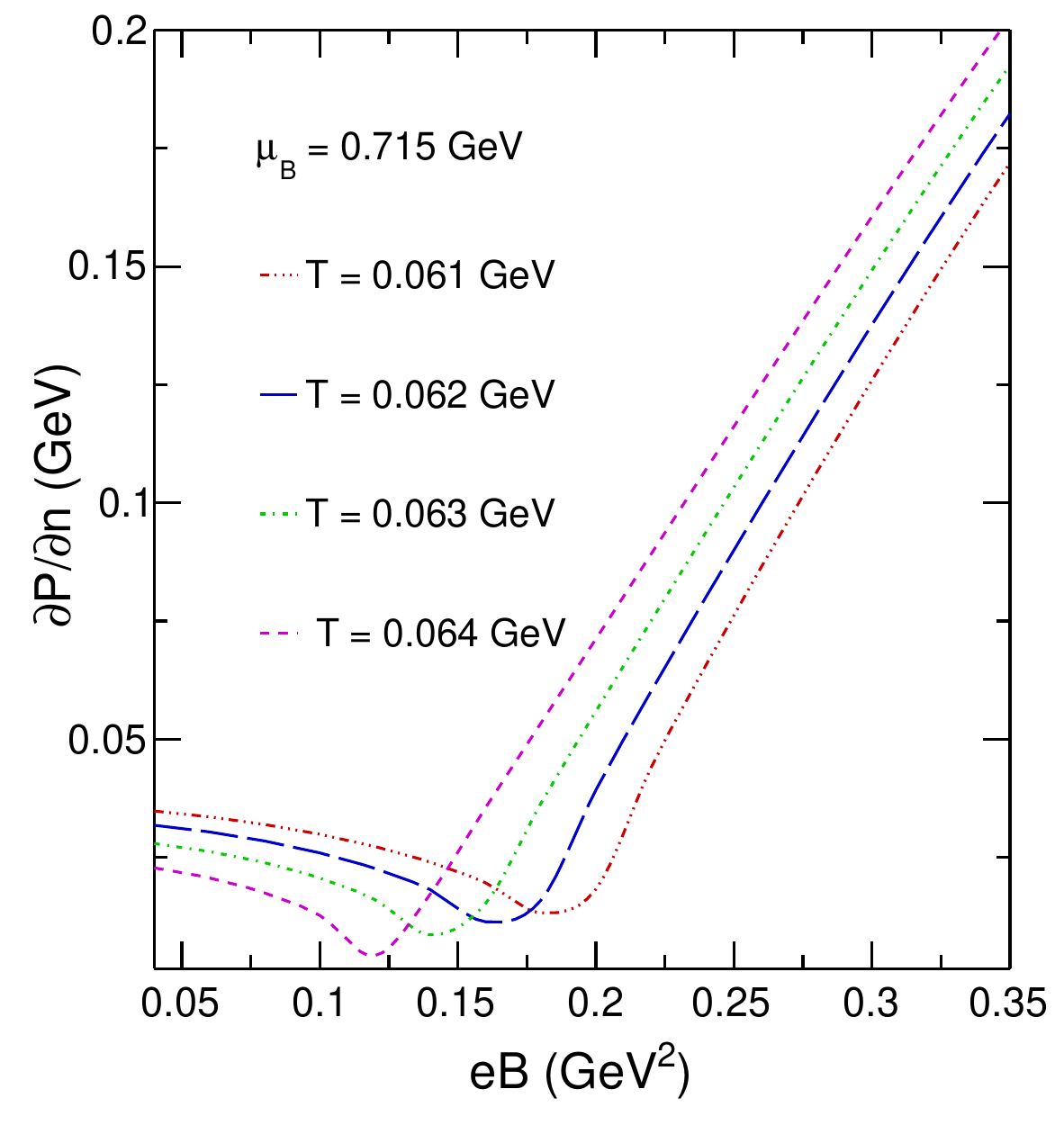}
    \caption{} \label{fig:4a}
  \end{subfigure}%
  \hspace*{1.5cm}   
\begin{subfigure}{0.35\textwidth}
\includegraphics[scale = 0.35]{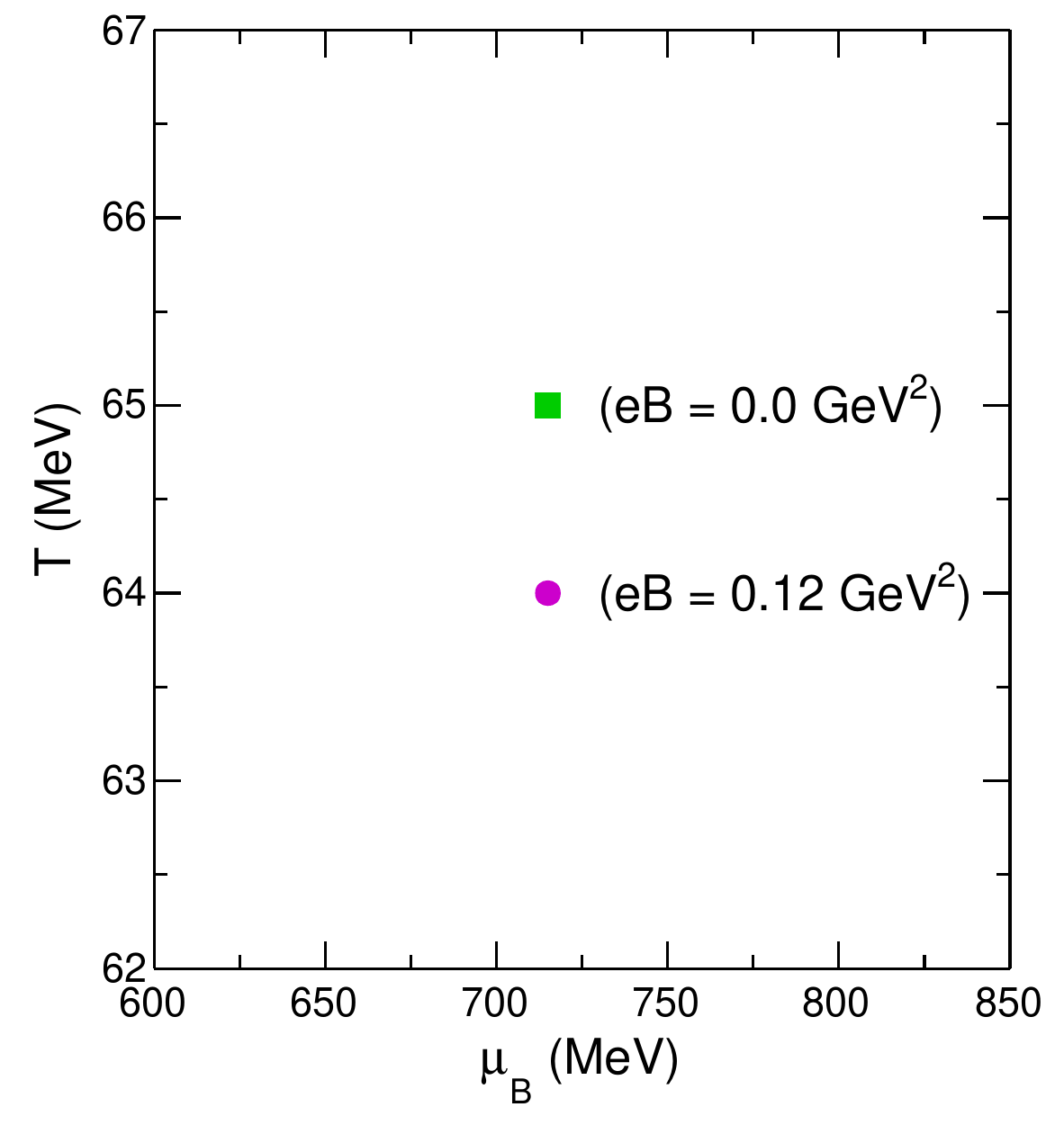}
 \caption{} \label{fig:4b}
  \end{subfigure}%

\caption{The left panel shows the variation of $(\partial P/ \partial n)_{T}$ as a function of magnetic field $eB$. The right panel shows the critical point of the liquid-gas phase transition in the QCD phase diagram in the presence of a magnetic field. }
\label{fig5}
\end{figure*}

We estimate the electric charge number susceptibility in HRG and VDWHRG models using Eq.~(\ref{eq23}). Figure~\ref{fig:4} shows the temperature dependence of electrical susceptibility for different values of the magnetic field. One observes that electric charge number susceptibility increases with an increase in temperature. With a higher magnetic field, the electric charge number susceptibility is found to be suppressed at lower temperatures, and it starts to increase beyond a certain value of temperature. This limiting temperature was found to decrease with an increase in the magnetic field. This is because the dominant contribution to susceptibility comes from spin-0 particles ($\pi^{\pm}, k^{\pm}$, etc.), and in the presence of a magnetic field, these particles are suppressed and do not contribute to susceptibility. As a result, the electric susceptibility decreases at low temperatures with an increase in the magnetic field. However, as temperature increases, the higher-spin nonzero resonance particles ($ \rho^{\pm}, k^{*\pm}, \Delta$, etc.) start contributing to susceptibility, and hence susceptibility is found to increase with the magnetic field at a higher temperature.

The ideal HRG model accounts only for the hadronic degrees of freedom without any phase transition to QGP. However, the inclusion of attractive and repulsive interaction through the VDWHRG model allows us to study the liquid-gas phase transition in the hadronic phase. In the literature, there are investigations of this liquid-gas phase transition in the VDWHRG model in the T-$\mu_{B}$ plane \cite{Samanta:2017yhh, Vovchenko:2015vxa, Sarkar:2018mbk}. With different interaction parameters $a$ and $b$, the critical point of the phase transition is found to be different. Here, we explore the effect of the magnetic field on this critical point and study the liquid-gas phase transition in the $T-\mu_{B}-eB$ plane using the VDWHRG model. In this analysis, we use the same van der Waals parameters as used in Ref. \cite{Sarkar:2018mbk}, where the authors observed the critical point around $T \approx 65 $ MeV, and $\mu_{B} \approx 715 $ MeV. Taking the same baryochemical potential, we explore the effect of the magnetic field to see its effect on the critical temperature. Figure ~\ref{fig5}(a) shows the variation of the $(\partial P/ \partial n)_{T}$ with $eB$ for the same chemical potential, $\mu_{B} = 715$ MeV. Each curve is for different temperatures taken for the calculation. One can observe that the $(\partial P/ \partial n)_{T}$ becomes zero at $T = 64$ MeV and $\mu_B = $715 MeV for $eB = 0.12$ GeV$^{2}$. This marks the critical temperature below which the number density varies discontinuously, showing the first-order liquid-gas phase transition. To demonstrate the role of the magnetic field on the critical point, we plot the critical points in the $T-\mu_{B}$ plane in Fig.~\ref{fig5}(b). The green square marker shows the critical point in the absence of the magnetic field \cite{Sarkar:2018mbk}, whereas the magenta circle marker shows the critical point in the presence of a magnetic field. One can observe that in the presence of the magnetic field, the critical point shifts towards lower temperatures, i.e., at $T$ = 0.064 GeV, $\mu_{B}$ = 0.715 GeV and eB = 0.12 $\rm GeV^{2}$. This indicates that the magnetic field delays the liquid-gas phase transition. It is also important to note that the critical point now depends on three parameters, namely, temperature, $T$, baryochemical potential, $\mu_{B}$, and the magnitude of the magnetic field, $eB$. Hence one can in principle be able to study the three-dimensional variation of the critical point in the $T-\mu_{B}-eB$ plane.

\section{Summary}
\label{sum}

In this work, we explore the effect of a magnetic field on the thermodynamic properties of an interacting hadron resonance gas model at zero and finite chemical potential. The static finite magnetic field significantly affects pressure, energy density, trace anomaly, magnetization, and second-order conserved charge fluctuations such as electric and magnetic susceptibility. However, this effect is less significant on entropy density, specific heat, etc. We found that all thermodynamic quantities are suppressed because of interactions. The effect of higher baryon chemical potential on the thermodynamic variable is interesting. The magnetization, entropy density, specific heat, and speed of sound may indicate discontinuity behavior approaching a higher baryochemical potential, which suggests a phase transition in the VDWHRG model. A clear observation of diamagnetic to paramagnetic transitions happens in our study. The electrical susceptibility is found to be suppressed because of the magnetic field at lower temperatures, and it slowly increases at higher temperatures.  A possible liquid-gas phase transition is also explored in the presence of a finite magnetic field and baryochemical potential. \\

\section*{Acknowledgements}
B.S. and K.K.P. acknowledge the financial aid from CSIR and UGC, the Government of India, respectively. The authors gratefully acknowledge the DAE-DST, Government of India funding under the mega-science project "Indian Participation in
the ALICE experiment at CERN" bearing Project No.
SR/MF/PS-02/2021-IITI (E-37123). The authors would like to acknowledge some fruitful discussions with Girija Sankar Pradhan during the preparation of the manuscript.

\appendix

\section{Squared speed of sound}
\label{appe1}
The squared speed of sound is given by,
\begin{equation}
\label{eq101}
c_{s}^{2}(T,\mu, QB) = \bigg(\frac{\partial P}{\partial \epsilon}\bigg)_{s/n}.
\end{equation}

Using the variables $T$, $\mu$ and $QB$ this can be rewritten as

\begin{equation}
\label{eq102}
c_{s}^{2}(T,\mu, QB) =\frac{d P}{d \epsilon} = \frac{\frac{\partial P}{\partial T} + \frac{\partial P}{\partial \mu}\frac{d\mu}{dT}+\frac{\partial P}{\partial (QB)}\frac{d(QB)}{dT}}{\frac{\partial \epsilon}{\partial T} +\frac{\partial \epsilon}{\partial \mu}\frac{d\mu}{dT}+ \frac{\partial \epsilon}{\partial (QB)}\frac{d(QB)}{dT}}
\end{equation}

Number density ($n$) and entropy density ($s$) of a system is a function of $(T,\mu, QB)$
The first condition is keeping the ratio $(s/n)$ constant. From the derivative, one obtains

\begin{equation}
\label{eq103}
d\bigg(\frac{s}{n}\bigg) = 0,
\end{equation}

which implies
\begin{equation}
\label{eq104}
 nds = sdn.
\end{equation}

Divide both sides by dT so that the above Eq. ~(\ref{eq4.1}) can be modified as,

\begin{equation}
\label{eq105}
n\bigg(\frac{ds}{dT}\bigg) = s\bigg(\frac{dn}{dT}\bigg).
\end{equation}

One can write $n (T,\mu, QB)$ and $s (T,\mu, QB)$ in the form of differential as,

\begin{equation}
\label{eq106}
dn = \frac{\partial n}{\partial T}dT + \frac{\partial n}{\partial \mu}d\mu+\frac{\partial n}{\partial (QB)}d (QB)
\end{equation} 
So if we divide both sides of the Eq. ~(\ref{eq106}) by dT, then we have,

\begin{equation}
\label{eq107}
\frac{dn}{dT} = \frac{\partial n}{\partial T} + \frac{\partial n}{\partial \mu}\frac{d\mu}{dT}+\frac{\partial n}{\partial (QB)}\frac{d (QB)}{dT}
\end{equation} 

Similarly for $s (T,\mu, QB)$ we can write,

\begin{equation}
\label{eq108}
ds = \frac{\partial s}{\partial T}dT + \frac{\partial s}{\partial \mu}d\mu+\frac{\partial s}{\partial (QB)}d (QB)
\end{equation} 

\begin{equation}
\label{eq109}
\frac{ds}{dT} = \frac{\partial s}{\partial T} + \frac{\partial s}{\partial \mu}\frac{d\mu}{dT}+\frac{\partial s}{\partial (QB)}\frac{d (QB)}{dT}
\end{equation} 

Substituting Eq.~(\ref{eq107}) and Eq. ~(\ref{eq109}) in Eq. ~(\ref{eq105}) we get,
\begin{widetext}
    \begin{equation}
\label{eq101}
n\bigg(\frac{\partial s}{\partial T} + \frac{\partial s}{\partial \mu}\frac{d\mu}{dT}+\frac{\partial s}{\partial (QB)}\frac{d (QB)}{dT}\bigg) = s\bigg(\frac{\partial n}{\partial T} + \frac{\partial n}{\partial \mu}\frac{d\mu}{dT}+\frac{\partial n}{\partial (QB)}\frac{d (QB)}{dT}\bigg).
\end{equation}

\begin{eqnarray}
\label{eq111}
\frac {d\mu}{dT}\bigg(n \frac{\partial s}{\partial \mu}-s\frac{\partial n}{\partial \mu}\bigg) = s\bigg(\frac{\partial n}{\partial T}+\frac{\partial n}{\partial (QB)}\frac{d(QB)}{ dT}\bigg) - n\bigg(\frac{\partial s}{\partial T}+\frac{\partial s}{\partial (QB)}\frac{d(QB)}{dT}\bigg)\nonumber\\
\Rightarrow \frac {d\mu}{dT} = \frac{\bigg(s\frac{\partial n}{\partial T}- n\frac{\partial s}{\partial T}\bigg) + \bigg(s \frac{\partial n}{\partial (QB)}- n\frac{\partial s}{\partial (QB)} \bigg)\frac{d(QB)}{dT}}{n\frac{\partial s}{\partial \mu}-s\frac{\partial n}{\partial \mu}}
\end{eqnarray}

Similarly, one can evaluate $\frac{d(QB)}{dT}$ from Eq.~(\ref{eq101}) as follows,

\begin{equation}
\label{eq121}
\frac {d(QB)}{dT} = \frac{\bigg(s\frac{\partial n}{\partial T}- n\frac{\partial s}{\partial T}\bigg) + \bigg(s \frac{\partial n}{\partial \mu}- n\frac{\partial s}{\partial \mu} \bigg)\frac{d\mu}{dT}}{n\frac{\partial s}{\partial(QB)}-s\frac{\partial n}{\partial(QB)}}
\end{equation}

For a finite baryon chemical potential and finite external magnetic field, the above two transcendental equations can be solved numerically to find the speed of sound of the system. 

 \section{Magnetic susceptibility ($\chi_{M}^{2}$)}
 \label{appe2}
 In this paper, the magnetic susceptibility is calculated using Eq. (\ref{eq22}),

 \begin{align} 
     \chi_{M,i}^{2} = \frac{\pm g_i}{2\pi^2} \int_{0}^{\infty} dp_z \Big[Q_i B \Big( -\frac{(k+\frac{1}{2} -s_z )^2}{(E^{z})^2T[\exp(\frac{E^{z}_i -\mu_i}{T}) \pm 1 ]^{2}} \pm \frac{(k+\frac{1}{2} -s_z )^2}{(E^{z})^2T[\exp(\frac{E^{z}_i -\mu_i}{T}) \pm 1 ]^{}} \pm  \frac{(k+\frac{1}{2} -s_z )^2}{(E^{z})^{3}[\exp(\frac{E^{z}_i -\mu_i}{T}) \pm 1 ]^{}} \Big)  \nonumber \\  \mp \frac{2(k+\frac{1}{2} -s_z )}{E^{z}[\exp(\frac{E^{z}_i -\mu_i}{T}) \pm 1 ]} \Big].
     \label{eq55}  
 \end{align}

\section{Electrical susceptibility ($\chi_{Q}^{2}$)}
\label{appe3}
 The electrical susceptibility is calculated using Eq.~(\ref{eq23}),
 
\begin{align}   
     \chi_{Q,i}^{2} = \frac{g_{i}Q_{i}^{3}B}{2\pi^{2}T^{3}} \int_{0}^{\infty} dp_z \frac{\exp(\frac{E^{z}_i -\mu_i}{T})}{[\exp(\frac{E^{z}_i -\mu_i}{T}) \pm 1 ]^{2}}.
 \end{align}
 
\section{Vacuum contribution for Magnetization ($ \Delta M$)}
\label{appe4}
The explicit form of vacuum contribution for magnetization is obtained using Eq.~(\ref{eq54}).

For spin-0 particles,
\begin{align}
    \Delta \mathcal{M}_{vac}^r (S=0,B) =  \frac{\partial (\Delta P_{\text{vac}}(S =0,B))}{\partial (eB)}.
\end{align}
On simplifying,
\begin{align}
    \Delta \mathcal{M}_{vac}^r (S=0,B) =& \frac{|Q|B}{8\pi^2} \Big[\frac{x}{12(x+1/2)} + x^2ln(x+1/2) - \frac{(x+1/2)^{-2}}{360} \Big(\frac{x}{x+1/2} -1 \Big) -  \frac{(1+lnx)}{12} \nonumber \\ & -  2\Big(\frac{1}{12} -\frac{x+1/2}{2} +\frac{(x+1/2)^2}{2} \Big) ln(x+1/2) \Big]. 
\end{align}

For spin-1/2 particles,

\begin{align}
    \Delta \mathcal{M}_{vac}^r (S=1/2,B) =  \frac{\partial (\Delta P_{\text{vac}}(S =1/2,B))}{\partial (eB)}, 
\end{align}
\begin{align}
    \Delta \mathcal{M}_{vac}^r (S=1/2,B) =  \frac{-|Q|B }{720\pi^2x^2}.  
\end{align}
Similarly, for spin-1 particles one can write,
\begin{align}
    \Delta \mathcal{M}_{vac}^r (S=1/2,B) =  \frac{\partial (\Delta P_{\text{vac}}(S =1,B))}{\partial (eB)}, 
\end{align}

\begin{align}
    \Delta \mathcal{M}_{vac}^r (S=1,B) = & \frac{3|Q|B}{8\pi^2} \Big[\frac{x}{12(x-1/2)} + x^2ln(x-1/2) - \frac{(x-1/2)^{-2}}{360} \Big(\frac{x}{x-1/2} -1 \Big) - 2\Big(\frac{1}{12} -\frac{x-1/2}{2} +\frac{(x-1/2)^2}{2} \Big) \nonumber \\ & ln(x-1/2)  - \frac{(x+1)ln(x+1/2)}{3}  +  \frac{(2-5x)ln(x-1/2)}{3}+ \frac{(3+7lnx)}{12} \Big],      
\end{align}

where, $ x =  \frac{m^2}{2|Q|B}$. Now, the total vacuum contribution consists of the contribution from the spin-0, spin-1/2 and spin-1 particles. So, we finally get,
\begin{align}
\Delta \mathcal{M}_{vac}^r = \Delta \mathcal{M}_{vac}^r(S=0,B) + \Delta \mathcal{M}_{vac}^r(S=1/2,B) +\Delta \mathcal{M}_{vac}^r(S=1,B).     
\end{align}
\end{widetext}

\vspace{10.0025cm}

\begin{thebibliography}{}
{
\bibitem{Borsanyi:2013bia}
S.~Borsanyi, Z.~Fodor, C.~Hoelbling, S.~D.~Katz, S.~Krieg and K.~K.~Szabo,
Phys. Lett. B \textbf{730}, 99 (2014).

\bibitem{HotQCD:2014kol}
A.~Bazavov \textit{et al.} [HotQCD],
Phys. Rev. D \textbf{90}, 094503 (2014).

\bibitem{Bellwied:2013cta}
R.~Bellwied, S.~Borsanyi, Z.~Fodor, S.~D.~Katz and C.~Ratti,
Phys. Rev. Lett. \textbf{111}, 202302 (2013).

\bibitem{HotQCD:2012fhj}
A.~Bazavov \textit{et al.} [HotQCD],
Phys. Rev. D \textbf{86}, 034509 (2012).

\bibitem{Bellwied:2017ttj}
R.~Bellwied, S.~Borsanyi, Z.~Fodor, J.~Gunther, K.~H.~Kampert, S.~D.~Katz, T.~Kawanai, T.~G.~Kovacs, S.~W.~Mages and A.~Pasztor, \textit{et al.}
Nucl. Phys. A \textbf{967}, 732 (2017).

\bibitem{Borsanyi:2010cj}
S.~Borsanyi, G.~Endrodi, Z.~Fodor, A.~Jakovac, S.~D.~Katz, S.~Krieg, C.~Ratti and K.~K.~Szabo,
JHEP \textbf{11}, 077 (2010).

\bibitem{Karsch:2003vd}
F.~Karsch, K.~Redlich and A.~Tawfik,
Eur. Phys. J. C \textbf{29}, 549 (2003).

\bibitem{Huovinen:2009yb}
P.~Huovinen and P.~Petreczky,
Nucl. Phys. A \textbf{837}, 26 (2010).

\bibitem{Karsch:2003zq}
F.~Karsch, K.~Redlich and A.~Tawfik,
Phys. Lett. B \textbf{571}, 67 (2003).

\bibitem{Tawfik:2004sw}
A.~Tawfik,
Phys. Rev. D \textbf{71}, 054502 (2005).

\bibitem{Allton:2005gk}
C.~R.~Allton, M.~Doring, S.~Ejiri, S.~J.~Hands, O.~Kaczmarek, F.~Karsch, E.~Laermann and K.~Redlich,
Phys. Rev. D \textbf{71}, 054508 (2005).
\bibitem{Borsanyi:2012cr}
S.~Borsanyi, G.~Endrodi, Z.~Fodor, S.~D.~Katz, S.~Krieg, C.~Ratti and K.~K.~Szabo,
JHEP \textbf{08}, 053 (2012).

\bibitem{Braun:2004}
P. Braun-Munzinger, K. Redlich, and J. Stachel, in Quark Gluon
Plasma 3, edited by R. C. Hwa and X. N. Wang (World Scientific,
Singapore, 2004).


\bibitem{Dashen:1969ep}
R.~Dashen, S.~K.~Ma and H.~J.~Bernstein,
Phys. Rev. \textbf{187}, 345 (1969).

\bibitem{Venugopalan:1992hy}
R.~Venugopalan and M.~Prakash,
Nucl. Phys. A \textbf{546}, 718 (1992).

\bibitem{Bhattacharyya:2015pra}
A.~Bhattacharyya, S.~K.~Ghosh, R.~Ray and S.~Samanta,
EPL \textbf{115}, 62003 (2016).


\bibitem{Hagedorn:1980kb}
R.~Hagedorn and J.~Rafelski,
Phys. Lett. B \textbf{97}, 136 (1980).

\bibitem{Rischke:1991ke}
D.~H.~Rischke, M.~I.~Gorenstein, H.~Stoecker and W.~Greiner,
Z. Phys. C \textbf{51}, 485 (1991).

\bibitem{Cleymans:1992jz}
J.~Cleymans, M.~I.~Gorenstein, J.~Stalnacke and E.~Suhonen,
Phys. Scripta \textbf{48}, 277 (1993).

\bibitem{Yen:1997rv}
G.~D.~Yen, M.~I.~Gorenstein, W.~Greiner and S.~N.~Yang,
Phys. Rev. C \textbf{56}, 2210 (1997).
\bibitem{Fu:2012zzc}
J.~Fu,
Phys. Rev. C \textbf{85}, 064905 (2012).

\bibitem{Fu:2013gga}
J.~Fu,
Phys. Lett. B \textbf{722}, 144 (2013).

\bibitem{Andronic:2012ut}
A.~Andronic, P.~Braun-Munzinger, J.~Stachel and M.~Winn,
Phys. Lett. B \textbf{718}, 80 (2012).
\bibitem{Singh:1991np}
C.~P.~Singh, B.~K.~Patra and K.~K.~Singh,
Phys. Lett. B \textbf{387}, 680 (1996).

\bibitem{Gorenstein:2007mw}
M.~I.~Gorenstein, M.~Hauer and O.~N.~Moroz,
Phys. Rev. C \textbf{77},  024911 (2008).

\bibitem{Tawfik:2013eua}
A.~Tawfik,
Phys. Rev. C \textbf{88}, 035203 (2013).

\bibitem{Begun:2012rf}
V.~V.~Begun, M.~Gazdzicki and M.~I.~Gorenstein,
Phys. Rev. C \textbf{88}, 024902 (2013).

\bibitem{Bhattacharyya:2013oya}
A.~Bhattacharyya, S.~Das, S.~K.~Ghosh, R.~Ray and S.~Samanta,
Phys. Rev. C \textbf{90}, 034909 (2014).




\bibitem{Bugaev:2000wz}
K.~A.~Bugaev, M.~I.~Gorenstein, H.~Stoecker and W.~Greiner,
Phys. Lett. B \textbf{485}, 121 (2000).

\bibitem{Sagun:2013moa}
V.~V.~Sagun, A.~I.~Ivanytskyi, K.~A.~Bugaev and I.~N.~Mishustin,
Nucl. Phys. A \textbf{924}, 24 (2014).

\bibitem{Pal:2021qav}
S.~Pal, G.~Kadam and A.~Bhattacharyya,
Nucl. Phys. A \textbf{1023}, 122464 (2022).
\bibitem{Kadam:2019peo}
G.~Kadam and H.~Mishra,
Phys. Rev. D \textbf{100}, 074015 (2019).
\bibitem{Anchishkin:2014hfa}
D.~Anchishkin and V.~Vovchenko,
J. Phys. G \textbf{42}, 105102 (2015).

\bibitem{Pal:2020ink}
S.~Pal, A.~Bhattacharyya and R.~Ray,
Nucl. Phys. A \textbf{1010}, 122177 (2021).


\bibitem{Zhang:2019uct}
H.~X.~Zhang, J.~W.~Kang and B.~W.~Zhang,
Phys. Rev. D \textbf{101}, 114033 (2020).

\bibitem{Samanta:2017yhh}
S.~Samanta and B.~Mohanty,
Phys. Rev. C \textbf{97}, 015201 (2018).

\bibitem{Vovchenko:2015vxa}
V.~Vovchenko, D.~V.~Anchishkin and M.~I.~Gorenstein,
Phys. Rev. C \textbf{91}, 064314 (2015).

\bibitem{Vovchenko:2016rkn}
V.~Vovchenko, M.~I.~Gorenstein and H.~Stoecker,
Phys. Rev. Lett. \textbf{118}, 182301 (2017).

\bibitem{Sarkar:2018mbk}
N.~Sarkar and P.~Ghosh,
Phys. Rev. C \textbf{98}, 014907 (2018).

\bibitem{Pradhan:2022gbm}
K.~K.~Pradhan, D.~Sahu, R.~Scaria and R.~Sahoo, Phys. Rev. C \textbf{107}, 014910 (2023).

\bibitem{Skokov:2009qp}
V.~Skokov, A.~Y.~Illarionov and V.~Toneev,
Int. J. Mod. Phys. A \textbf{24}, 5925 (2009).

\bibitem{Tawfik:2016lih}
A.~N.~Tawfik, A.~M.~Diab, N.~Ezzelarab and A.~G.~Shalaby,
Adv. High Energy Phys. \textbf{2016}, 1381479  (2016).

\bibitem{Bzdak:2011yy}
A.~Bzdak and V.~Skokov,
Phys. Lett. B \textbf{710}, 171 (2012).

\bibitem{Deng:2012pc}
W.~T.~Deng and X.~G.~Huang,
Phys. Rev. C \textbf{85}, 044907 (2012).

\bibitem{Duncan:1992hi}
R.~C.~Duncan and C.~Thompson,
Astrophys. J. Lett. \textbf{392}, L9 (1992). 
\bibitem{Dey:2002pc}
P.~Dey, A.~Bhattacharyya and D.~Bandyopadhyay,
J. Phys. G \textbf{28}, 2179 (2002).


\bibitem{Vachaspati:1991nm}
T.~Vachaspati,
Phys. Lett. B \textbf{265}, 258 (1991).

\bibitem{Bhatt:2015ewa}
J.~R.~Bhatt and A.~K.~Pandey,
Phys. Rev. D \textbf{94}, 043536 (2016).


\bibitem{Kharzeev:2007jp}
D.~E.~Kharzeev, L.~D.~McLerran and H.~J.~Warringa,
Nucl. Phys. A \textbf{803}, 227 (2008).

\bibitem{Fukushima:2008xe}
K.~Fukushima, D.~E.~Kharzeev and H.~J.~Warringa,
Phys. Rev. D \textbf{78}, 074033 (2008).
\bibitem{Bali:2011qj}
G.~S.~Bali, F.~Bruckmann, G.~Endrodi, Z.~Fodor, S.~D.~Katz, S.~Krieg, A.~Schafer and K.~K.~Szabo,
JHEP \textbf{02}, 044 (2012).

\bibitem{Shovkovy:2012zn}
I.~A.~Shovkovy,
Lect. Notes Phys. \textbf{871} 13 (2013).

\bibitem{Preis:2010cq}
F.~Preis, A.~Rebhan and A.~Schmitt,
JHEP \textbf{03},  033 (2011).
\bibitem{Bruckmann:2013oba}
F.~Bruckmann, G.~Endrodi and T.~G.~Kovacs,
JHEP \textbf{04}, 112 (2013).

\bibitem{Bali:2012jv}
G.~S.~Bali, F.~Bruckmann, M.~Constantinou, M.~Costa, G.~Endrodi, S.~D.~Katz, H.~Panagopoulos and A.~Schafer,
Phys. Rev. D \textbf{86}, 094512 (2012).

\bibitem{Bali:2020bcn}
G. S. Bali, G. Endr\H{o}di, and S. Piemonte,
JHEP \textbf{07}, 183 (2020).
\bibitem{Lin:2022ied}
F.~Lin, K.~Xu and M.~Huang,
Phys. Rev. D \textbf{106}, 016005 (2022). 
\bibitem{Xu:2020yag}
K.~Xu, J.~Chao and M.~Huang,
Phys. Rev. D \textbf{103}, 076015 (2021).
\bibitem{Bonati:2013vba}
C.~Bonati, M.~D'Elia, M.~Mariti, F.~Negro and F.~Sanfilippo,
Phys. Rev. D \textbf{89}, 054506 (2014).
\bibitem{Bonati:2013lca}
C.~Bonati, M.~D'Elia, M.~Mariti, F.~Negro and F.~Sanfilippo,
Phys. Rev. Lett. \textbf{111}, 182001 (2013).

\bibitem{Bali:2014kia}
G.~S.~Bali, F.~Bruckmann, G.~Endr\"odi, S.~D.~Katz and A.~Sch\"afer,
JHEP \textbf{08}, 177 (2014).

\bibitem{Chaudhuri:2022oru}
N.~Chaudhuri, S.~Ghosh, P.~Roy and S.~Sarkar,
Phys. Rev. D \textbf{106}, 056020 (2022).

\bibitem{Tawfik:2016gye}
A.~N.~Tawfik, A.~M.~Diab and M.~T.~Hussein,
J. Phys. G \textbf{45}, 055008 (2018).

\bibitem{Endrodi:2013cs}
G.~Endr\"odi,
JHEP \textbf{04}, 023 (2013).

\bibitem{Kadam:2019rzo}
G.~Kadam, S.~Pal and A.~Bhattacharyya,
J. Phys. G \textbf{47}, 125106 (2020)

\bibitem{Pradhan:2021vtp}
G.~S.~Pradhan, D.~Sahu, S.~Deb and R.~Sahoo,
J. Phys. G \textbf{50}, 055104 (2023).



\bibitem{Landau}
L. Landau and E. Lifshits, Quantum mechanics: non-relativistic theory, Course of theoretical
physics volume 3, Pergamon Press, U.K. (1977).



\bibitem{Chakrabarty:1996te}
S.~Chakrabarty,
Phys. Rev. D \textbf{54}, 1306 (1996).

\bibitem{Fraga:2008qn}
E.~S.~Fraga and A.~J.~Mizher,
Phys. Rev. D \textbf{78}, 025016 (2008).

\bibitem{Kittel}
 C. Kittel, Elementary statistical physics, Dover Books on Physics Series, Dover Publications,
U.S.A. (2004).
\bibitem{Stanley}
H. Stanley, Introduction to phase transitions and critical phenomena, The International
Series of Monographs on Physics Series, Oxford University Press, Oxford U.K. (1987).

\bibitem{Ferrara:1992yc}
S.~Ferrara, M.~Porrati and V.~L.~Telegdi,
Phys. Rev. D \textbf{46}, 3529 (1992).

\bibitem{Goldman:1972vc}
J.~T.~Goldman, W.~Y.~Tsai and A.~Yildiz,
Phys. Rev. D \textbf{5}, 1926 (1972).

\bibitem{Tsai:1971zma}
W.~Y.~Tsai and A.~Yildiz,
Phys. Rev. D \textbf{4}, 3643 (1971).

\bibitem{Vovchenko:2015pya}
V.~Vovchenko, D.~V.~Anchishkin, M.~I.~Gorenstein and R.~V.~Poberezhnyuk,
Phys. Rev. C \textbf{92}, 054901 (2015). 

\bibitem{Menezes:2008qt}
D.~P.~Menezes, M.~Benghi Pinto, S.~S.~Avancini, A.~Perez Martinez and C.~Providencia,
Phys. Rev. C \textbf{79}, 035807 (2009).


\bibitem{Peskin:1995}
M. Peskin and D. Schroeder, An introduction to quantum field theory, Westview Press,
U.S.A. (1995).
\bibitem{Ramond:2001}
Ramond P 2001 Field Theory: A Modern Primer (CO, USA: Westview Press)

\bibitem{Digital:mathematica}
Digital library of mathematical functions, release date 2012-03-23, National Institute of
Standards and Technology, http://dlmf.nist.gov/.
\bibitem{Elizalde:1986}
Elizalde E 1986 An asymptotic expansion for the first derivative of the generalized Riemann zeta
function Math. Comp. \textbf{47} 347.

\bibitem{Schwinger:1951nm}
J.~S.~Schwinger,
Phys. Rev. \textbf{82}, 664 (1951).

\bibitem{Elmfors:1993bm}
P.~Elmfors, D.~Persson and B.~S.~Skagerstam,
Astropart. Phys. \textbf{2}, 299 (1994).

\bibitem{Andersen:2011ip}
J.~O.~Andersen and R.~Khan,
Phys. Rev. D \textbf{85}, 065026 (2012).

 \bibitem{PDG2016}
C.~Patrignani \textit{et al.} (Particle Data Group), Chin. Phys. C \textbf{40}, 100001 (2016).

\bibitem{Tawfik:2016sqd}
N.~A.~Tawfik, L.~I.~Abou-Salem, A.~G.~Shalaby, M.~Hanafy, A.~Sorin, O.~Rogachevsky and W.~Scheinast,
Eur. Phys. J. A \textbf{52}, 324 (2016).

\bibitem{Braun:2001sqd}
P. Braun-Munzinger, D. Magestro, K. Redlich, and J. Stachel,
Phys. Lett. B 518, 41 (2001).

\bibitem{Cleymans:2006sqd}
J. Cleymans, H. Oeschler, K. Redlich, and S. Wheaton, Phys.
Rev. C 73, 034905 (2006).
\bibitem{Khuntia:2019sqd}
A. Khuntia, S. K. Tiwari, P. Sharma, R. Sahoo, and T. K. Nayak,
Phys. Rev. C 100, 014910 (2019).


  
}
   

\end{thebibliography}
 \end{document}